\documentclass[10pt,twocolumn,letterpaper]{article}

\usepackage[pagenumbers]{cvpr} 

\newcommand{\acronym}[1]{SpeeDe3DGS}
\newcommand{\metrictableslow}[1]{{\colorbox[RGB]{230,160,160}{#1}}}
\newcommand{\metrictablebest}[1]{{\colorbox[RGB]{134,230,85}{#1}}}
\newcommand{\metrictablesecond}[1]{{\colorbox[RGB]{208,240,192}{#1}}}
\newcommand{\metrictableneural}[1]{{#1}}

\usepackage[utf8]{inputenc} 
\usepackage[T1]{fontenc}    
\usepackage{url}            
\usepackage{booktabs}       
\usepackage{amsfonts}       
\usepackage{nicefrac}       
\usepackage{microtype}      
\usepackage{xcolor}         
\usepackage{amsmath}
\newcommand{\Mat}{\boldsymbol}
\newcommand{\Set}{\mathcal}

\newcommand{\real}{\mathbb{R}}

\usepackage{graphicx}
\usepackage{multirow}
\usepackage{float}
\usepackage{placeins}
\usepackage{multicol}

\definecolor{cvprblue}{rgb}{0.21,0.49,0.74}
\usepackage[pagebackref,breaklinks,colorlinks,allcolors=cvprblue]{hyperref}

\def\paperID{*****} 
\def\confName{CVPR}
\def\confYear{2026}

\title{SpeeDe3DGS: Speedy Deformable 3D Gaussian Splatting\\with Temporal Pruning and Motion Grouping}

\author{%
  Allen Tu$^*$ 
  \hspace{2em}
  Haiyang Ying$^*$
  \hspace{2em}
  Alex Hanson
  \hspace{2em}
  Yonghan Lee
  \\
  Tom Goldstein
  \hspace{2em}
  Matthias Zwicker \\
    [0.35em]
    \textnormal{University of Maryland, College Park} \\
    [0.1em]
   \url{https://speede3dgs.github.io}  
  \vspace{-8mm}
}

\begin{document}
\twocolumn[{
\renewcommand\twocolumn[1][]{#1}
\maketitle

\begin{center}
    \includegraphics[width=\linewidth]{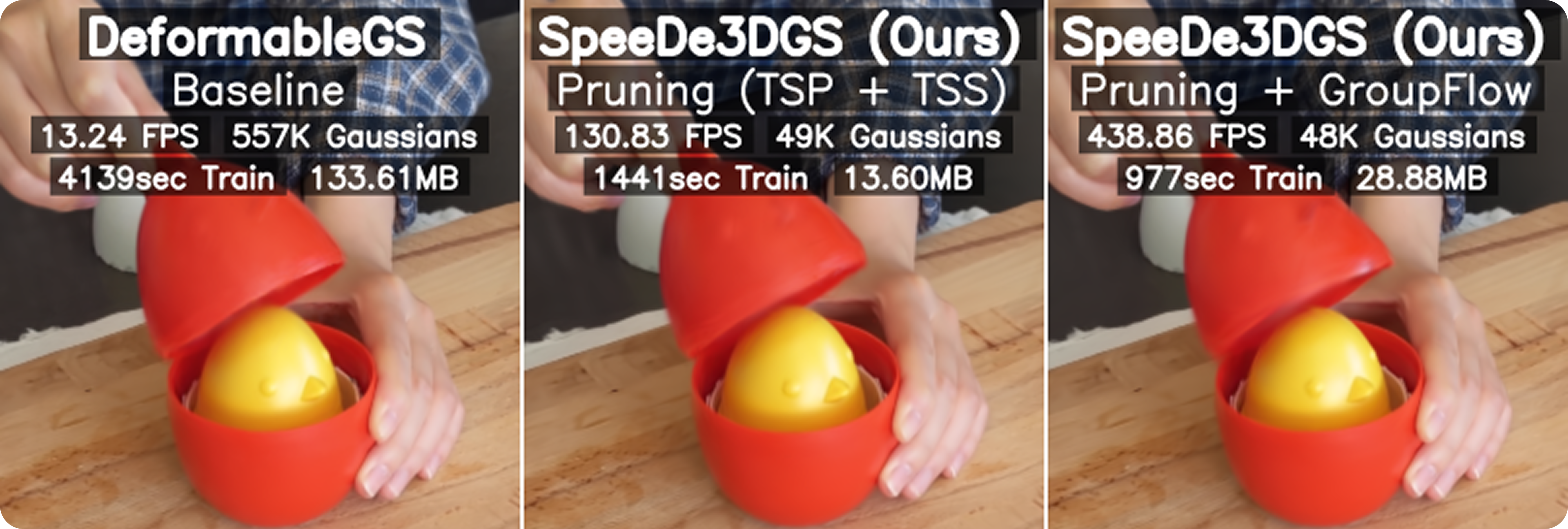}
    \vspace{-6mm}
    \captionsetup{type=figure}
    \caption{Our \acronym{} framework achieves $9.88\times$ faster rendering, $11.37\times$ fewer Gaussians, and $2.87\times$ shorter training on the HyperNeRF~\cite{park2021hypernerf} \emph{chicken} scene while preserving the image quality of DeformableGS~\cite{yang2023deformable3dgs} through Temporal Sensitivity Pruning (TSP) and Sampling (TSS). Applying our GroupFlow method on top of pruning accelerates rendering and training by $33.13\times$ and $4.24\times$, respectively.}
    \label{fig:teaser}
    \vspace{-0.5mm}
\end{center}
}]



\makeatletter
\renewcommand\@makefnmark{\hbox{* }} 
\makeatother
\setlength{\skip\footins}{7pt} 
\scriptsize
\footnotetext[1]{denotes equal contribution.}
\normalsize

\begin{abstract}
Dynamic extensions of 3D Gaussian Splatting (3DGS) achieve high-quality reconstructions through neural motion fields, but per-Gaussian neural inference makes these models computationally expensive. Building on DeformableGS, we introduce Speedy Deformable 3D Gaussian Splatting (SpeeDe3DGS), which bridges this efficiency-fidelity gap through three complementary modules: Temporal Sensitivity Pruning (TSP) removes low-impact Gaussians via temporally aggregated sensitivity analysis, Temporal Sensitivity Sampling (TSS) perturbs timestamps to suppress floaters and improve temporal coherence, and GroupFlow distills the learned deformation field into shared SE(3) transformations for efficient groupwise motion. On the 50 dynamic scenes in MonoDyGauBench, integrating TSP and TSS into DeformableGS accelerates rendering by 6.78$\times$ on average while maintaining neural-field fidelity and using 10$\times$ fewer primitives. Adding GroupFlow culminates in 13.71$\times$ faster rendering and 2.53$\times$ shorter training, surpassing all baselines in speed while preserving superior image quality. 
\end{abstract}
\vspace{-4mm}
\section{Introduction}
\label{sec:intro}

Neural Radiance Fields (NeRFs)~\cite{mildenhall2021nerf} have transformed novel-view synthesis by representing static scenes as continuous volumetric functions optimized through differentiable rendering.
While capable of photorealistic reconstruction, NeRFs rely on dense per-ray MLP inference, resulting in slow training and rendering.
3D Gaussian Splatting (3DGS)~\cite{kerbl3Dgaussians} replaces volumetric integration with differentiable rasterization of point-based Gaussian primitives, achieving real-time rendering with high fidelity and rapid convergence.
This efficiency has made 3DGS a foundation for modern approaches to 3D reconstruction.

Dynamic Gaussian Splatting extends this framework to time-varying scenes by coupling each Gaussian with a learned motion field.
Methods such as DeformableGS~\cite{yang2023deformable3dgs} and 4DGS~\cite{Wu2024CVPR4dgs} model temporal variation using neural deformation fields or spatiotemporal grids, enabling detailed dynamic reconstructions.
However, these approaches require per-Gaussian neural inference at every frame, leading to substantial computational cost. Monocular Dynamic Gaussian Splatting Benchmark (MonoDyGauBench)~\cite{liang2025monocular} presents the first systematic comparison across dynamic 3DGS variants, revealing that neural motion fields achieve the highest and most stable reconstruction quality but render several times slower than analytic or non-neural representations.

This efficiency–fidelity gap motivates methods that preserve the expressiveness of neural motion fields while approaching the speed of analytic models.
In this paper, we propose \textbf{Speedy Deformable 3D Gaussian Splatting (SpeeDe3DGS)}, a framework that bridges this gap by reducing redundant neural inference through three complementary modules.
\textbf{Temporal Sensitivity Pruning (TSP)} eliminates redundant or low-impact Gaussians to reduce deformation and rendering cost, while its companion, \textbf{Temporal Sensitivity Sampling (TSS)}, stabilizes pruning by probing temporally jittered timestamps to suppress floaters and improve temporal coherence.
Finally, \textbf{GroupFlow} distills the learned neural deformation field into shared rigid SE(3) transformations, grouping Gaussians with similar motion so that each cluster shares a single motion representation.
Together, these modules reduce redundant primitives, stabilize temporal pruning, and dramatically accelerate dynamic reconstruction without sacrificing neural-field fidelity.

On MonoDyGauBench, integrating TSP and TSS into \textbf{DeformableGS} accelerates rendering by $6.78\times$ and training by $2.18\times$, maintaining neural-field fidelity while achieving frame rates comparable to non-neural baselines and using $10\times$ fewer primitives.
Adding \textbf{GroupFlow} further increases rendering and training performance to $13.71\times$ and $2.53\times$, respectively, surpassing all baselines in speed while still outperforming every non-neural method in image quality.

In summary, we propose the following contributions:

\begin{enumerate}
\item \textbf{Temporal Sensitivity Pruning (TSP):} Temporally aggregated gradient-based pruning that reduces neural inference cost by removing low-impact Gaussians.
\item \textbf{Temporal Sensitivity Sampling (TSS):} Temporal perturbation during pruning to improve stability and suppress floaters in noisy or dynamic scenes.
\item \textbf{GroupFlow:} Grouped SE(3) motion distillation that replaces per-Gaussian deformation with shared rigid motion for efficient, temporally coherent rendering.
\end{enumerate}
\section{Related Work}
\label{sec:references}

Neural Radiance Fields (NeRF)~\cite{mildenhall2021nerf} represent scenes as continuous volumetric functions optimized via differentiable volume rendering.
Dynamic extensions incorporate time through explicit temporal modeling~\cite{du2021nerflow, gao2021dnvs, wang2021dctnerf} or deformation fields that map a canonical space to time-varying observations~\cite{pumarola2020dnerf, park2021nerfies, park2021hypernerf}.
Grid-based and factorized encodings~\cite{guo2022neural, liu2022devrf, peng2023mlpmap, fang2022tineuvox, fridovich2023k, cao2023hexplane, gao2023strivec, mihajlovic2024resfields, wang2023msth} improve efficiency, but rendering remains slow due to per-ray MLP evaluation.

3D Gaussian Splatting (3DGS)~\cite{kerbl3Dgaussians} replaces volumetric integration with differentiable rasterization of Gaussian primitives, enabling real-time rendering for static scenes.
Dynamic extensions incorporate explicit or implicit motion models to represent time-varying geometry.
Dy3DGS~\cite{luiten2023dynamic3dgs} fits 6-DoF trajectories to individual Gaussians, while DeformableGS~\cite{yang2023deformable3dgs} and 4DGS~\cite{Wu2024CVPR4dgs} learn neural deformation fields using MLPs or HexPlane grids~\cite{cao2023hexplane}.
RTGS~\cite{yang2023ICLR4dgs} encodes temporal variation implicitly within a 4D Gaussian distribution.
MonoDyGauBench~\cite{liang2025monocular} unifies these formulations and identifies five representative motion parameterizations -- EffGS~\cite{katsumata2023efficient}, STG~\cite{li2023spacetime}, DeformableGS~\cite{yang2023deformable3dgs}, 4DGS~\cite{Wu2024CVPR4dgs}, and RTGS~\cite{yang2023ICLR4dgs} -- each trading off reconstruction fidelity and efficiency through distinct models of motion.

\begin{figure*}[t]
    \centering
    \includegraphics[width=\linewidth]{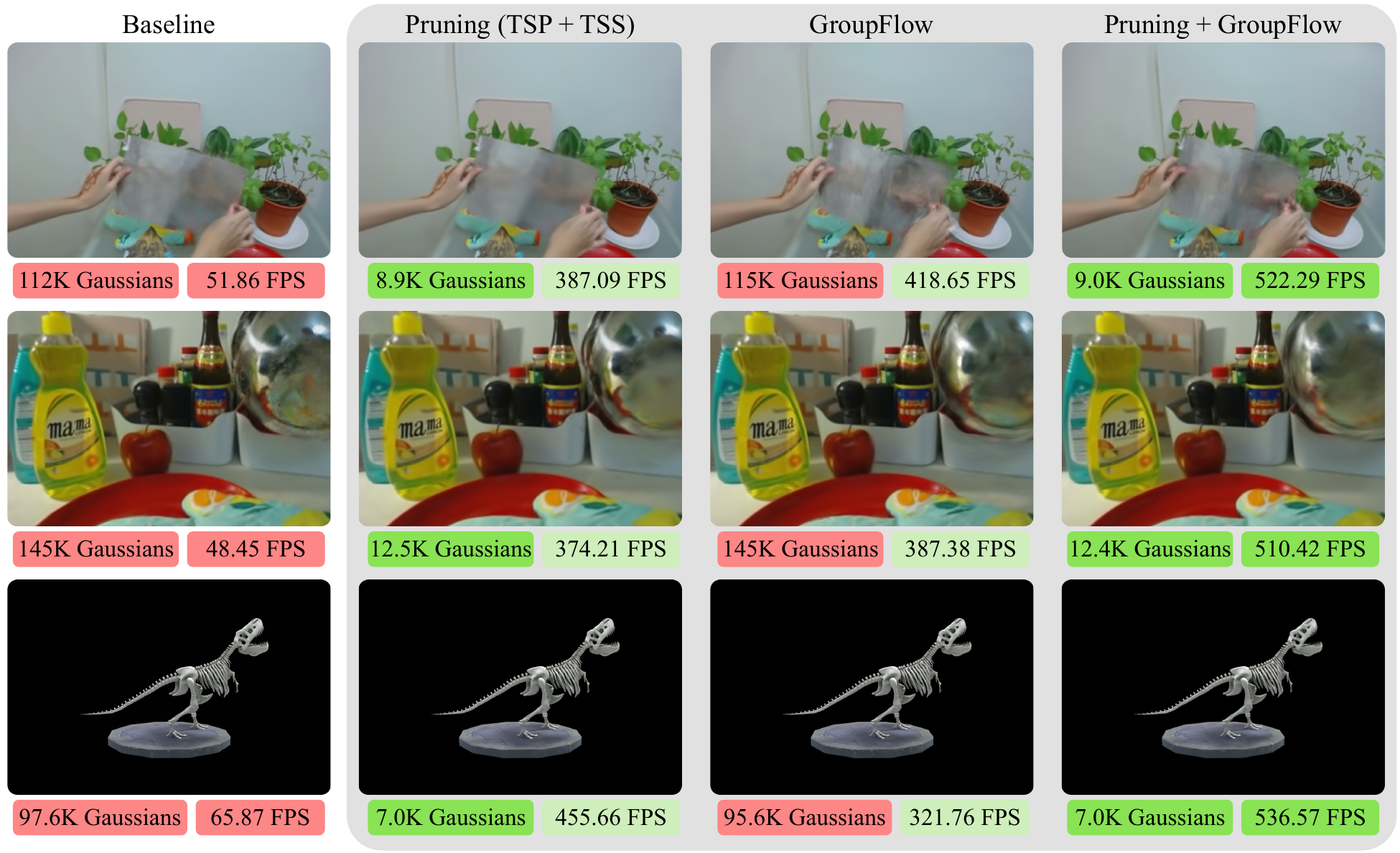}
    \vspace{-6mm}
    \caption{\textbf{Visual comparison of the baseline DeformableGS~\cite{yang2023deformable3dgs} and our \acronym{} methods.} Pruning (TSP + TSS) and GroupFlow deliver vastly faster results. Top: \emph{as} from NeRF-DS~\cite{yan2023nerf}. Middle: \emph{basin} from NeRF-DS. Bottom: \emph{trex} from D-NeRF~\cite{pumarola2020dnerf}.}
    \label{fig:methods_comparison}
    \vspace{-3mm}
\end{figure*}

\subsection{Gaussian Pruning}

Redundancy in Gaussian-based reconstructions has motivated extensive pruning research to improve efficiency and reduce memory overhead~\cite{fan2024lightgaussian, fang2024mini, niemeyer2024radsplat, HansonTuPUP3DGS, HansonSpeedy}.
Most approaches either learn per-Gaussian pruning masks~\cite{zhang2024lp3dgslearningprune3d, lee2024compact} or compute heuristic importance scores to remove low-contribution Gaussians~\cite{niemeyer2024radsplat, fan2024lightgaussian, fang2024mini, lin2024rtgs, girish2023eagles, ali2024elmgs, ali2024trimming, liu2024compgs, papantonakis2024reducing}.
Recent extensions adapt these strategies to dynamic scenes by incorporating temporal variation into pruning metrics~\cite{liu2024lgslightweight4dgaussian, javed2024temporallycompressed3dgaussian, zhang2024megamemoryefficient4dgaussian}.
PUP 3D-GS~\cite{HansonTuPUP3DGS} introduced a Hessian-based sensitivity score to quantify each Gaussian’s contribution to reconstruction, while Speedy-Splat~\cite{HansonSpeedy} refined this for pruning during training, achieving over $6\times$ faster rendering with minimal fidelity loss.
Building on these sensitivity-based principles, our approach generalizes gradient-based importance analysis to dynamic 3D Gaussian Splatting, enabling pruning that accounts for motion-dependent redundancy across time.

\subsection{Motion Analysis and Group-Based Modeling}
\label{sec:references:motion}

Another strategy for compressing dynamic scenes is to reduce temporal complexity by decomposing motion into static and dynamic regions.
Since most real-world scenes are largely static, several methods identify dynamic regions via motion cues~\cite{sabourgoli2024spotlesssplats, wang2025degauss, Zhou2024HUGS, xu2024das3r, wu2025swift4d} or cluster Gaussians using self-supervised or segmentation-based methods~\cite{wang2025degauss, xu2024das3r}, restricting expensive 4D modeling to dynamic areas.

However, static–dynamic decomposition alone is insufficient for complex non-rigid motion.
Group-based motion representations address this by jointly modeling correlated motion across sets of Gaussians.
SC-GS~\cite{huang2023scgs}, SPGaussian~\cite{wan2024superpoint}, and MoSca~\cite{lei2024mosca} model motion using control points with linear blend skinning or dual quaternion blending, often combined with MLP deformation. Methods such as DynMF~\cite{kratimenos2024dynmf}, Gaussian-Flow~\cite{lin2024gaussianflow}, and Shape of Motion~\cite{som2024} instead learn low-dimensional motion bases or flow fields.
While effective for improving visual quality, these methods require costly neural inference~\cite{huang2023scgs, wan2024superpoint}, complex deformation computation~\cite{lei2024mosca}, or guidance from 2D priors~\cite{lei2024mosca, som2024}.
In contrast, our approach distills neural motion into efficient grouped SE(3) transformations, clustering Gaussians by trajectory similarity to share rigid motion and reduce per-Gaussian inference while maintaining high fidelity.
\section{Background}
\subsection{3D Gaussian Splatting}
\label{sec:background:3dgs}

3D Gaussian Splatting (3DGS)~\cite{kerbl3Dgaussians} represents scenes as parametric, point-based models composed of 3D Gaussians.
Given a set of ground truth training images $\Set{I}_{gt} = \{\Mat{I}_i \in \real^{H \times W}\}_{i=1}^M$, the scene is initialized using Structure from Motion (SfM) to produce image pose estimates and a sparse point cloud that serves as the initial means for the 3D Gaussians. Each image is paired with its corresponding pose in $\Set{P}_{gt} = \{\phi_i \in \real^{3 \times 4}\}_{i=1}^M$ and used to optimize the scene.

Each 3D Gaussian primitive $\Set{G}_i$ is parameterized by three geometry parameters -- mean $\mu_i \in \real^3$, scale $s_i \in \real^3$, and rotation $r_i \in \real^4$ -- and two appearance parameters -- view-dependent spherical harmonic color $h_i \in \real^{16\times 3}$ and opacity $\sigma_i \in \real$.
The set of all parameters can be described as
\begin{equation}
\Set{G} = \{\Set{G}_i = \{\mu_i, s_i, r_i, h_i, \sigma_i\}\}_{i=1}^N,
\end{equation}
where $N$ is the number of Gaussians. Given camera pose $\phi$, the scene is rendered by projecting all Gaussians to image space and compositing them via alpha blending.
The value of the 2D projection of Gaussian $\Set{G}_i$ at pixel $p$ is given by:
\vspace{-2mm}
\begin{equation}
g_i = e^q, \quad q = -\frac{1}{2} (p - \mu_{i_{2D}}) \Mat{\Sigma_{i_{2D}}}^{-1} (p - \mu_{i_{2D}})^T,
\label{eq:2d_gaussian_scalar}
\end{equation}
where $\mu_{i_{2D}}$ is the projection of $\mu_i$ onto image space and $\Mat{\Sigma_{i_{2D}}}^{-1}$ is the inverse of the 2D covariance computed via the EWA Splatting approximation~\cite{zwicker2002ewa} of the perspectively projected 3D Gaussian. The model is optimized via stochastic gradient descent on an image reconstruction loss 
\begin{equation}
    L(\Set{G} | \phi, I_{gt} ) = ||I_{\Set{G}}(\phi)-I_{gt}||_1 + L_{\text{D-SSIM}}(I_\Set{G}(\phi),I_{gt}),
    \label{eq:gs_loss}
\end{equation}
where $I_{\Set{G}}(\phi)$ is the rendered image for pose $\phi$. During training, the scene is periodically densified by cloning and splitting uncertain Gaussians and pruned by removing large or transparent ones. While this formulation effectively models static geometry, it cannot capture the non-rigid or time-varying motion present in dynamic scenes.

\subsection{Dynamic Gaussian Splatting}
\label{sec:background:deformable}

Dynamic Gaussian Splatting extends 3DGS by coupling canonical Gaussians $\Set{G}$ with a deformation function $\mathcal{D}$ that predicts their evolution over time:
\begin{equation}
(\mu + \Delta\mu_t,\, r + \Delta r_t,\, s + \Delta s_t) = \mathcal{D}(\mu, r, s, t),
\label{eq:deformation}
\end{equation}
where $\mu$, $r$, and $s$ denote the canonical mean, rotation, and scale, and $(\Delta\mu_t, \Delta r_t, \Delta s_t)$ are time-dependent offsets.
At each timestep $t$, the deformed set $\Set{G}_t$ is rendered by the same differentiable rasterizer as static 3DGS, with each training image paired with its corresponding pose $\phi$ and timestamp $t$.

Recent dynamic Gaussian Splatting methods differ in how they parameterize $\mathcal{D}$, balancing motion expressiveness against computational efficiency.
MonoDyGauBench~\cite{liang2025monocular} categorizes representative formulations into five major types:

\begin{enumerate}
    \item \textbf{EffGS}~\cite{katsumata2023efficient} models per-Gaussian motion using second-order Fourier and first-order polynomial bases.
    \item \textbf{SpaceTimeGaussians (STG)}~\cite{li2023spacetime} combines radial basis and polynomial bases to represent locally supported motion trajectories.
    \item \textbf{DeformableGS}~\cite{yang2023deformable3dgs} learns a shared MLP deformation field conditioned on spatial position and time.
    \item \textbf{4D Gaussian Splatting (4DGS)}~\cite{Wu2024CVPR4dgs} replaces the MLP with a factorized HexPlane grid~\cite{cao2023hexplane} for more efficient spatiotemporal interpolation.
    \item \textbf{RTGS}~\cite{yang2023ICLR4dgs} is the only approach without an explicit motion model, encoding temporal variation implicitly within a 4D Gaussian distribution.
\end{enumerate}

Among these, \textbf{DeformableGS} and \textbf{4DGS} represent the canonical \emph{neural motion field} formulations.
They consistently achieve the highest reconstruction quality among existing methods but at substantial computational cost, as their continuous deformation fields require neural inference for every Gaussian at every frame.
This tradeoff between fidelity and efficiency motivates techniques that preserve high visual quality while reducing per-Gaussian inference overhead.
\section{Method}
\label{sec:method}

Dynamic Gaussian Splatting methods that rely on \emph{neural motion fields} effectively capture complex non-rigid motion, but their high computational cost limits real-time performance.
To overcome this bottleneck, we introduce two complementary strategies that significantly improve efficiency while maintaining high visual fidelity: \textbf{Temporal Sensitivity Pruning (TSP)} and \textbf{GroupFlow}.
TSP eliminates redundant or unstable Gaussians through temporally aggregated sensitivity analysis, while its submodule, \textbf{Temporal Sensitivity Sampling (TSS)}, introduces temporally jittered sampling to stabilize pruning and suppress floating artifacts.
GroupFlow further distills the learned neural motion field into grouped SE(3) transformations that share rigid motion within coherent regions.
We integrate these components into \textbf{Speedy Deformable 3D Gaussian Splatting (SpeeDe3DGS)}, a unified pipeline that reduces redundant primitives, stabilizes temporal pruning, and lowers deformation inference cost -- achieving neural reconstruction quality at rendering speeds comparable to non-neural motion representations.

\begin{figure*}[t]
  \includegraphics[width=\linewidth]{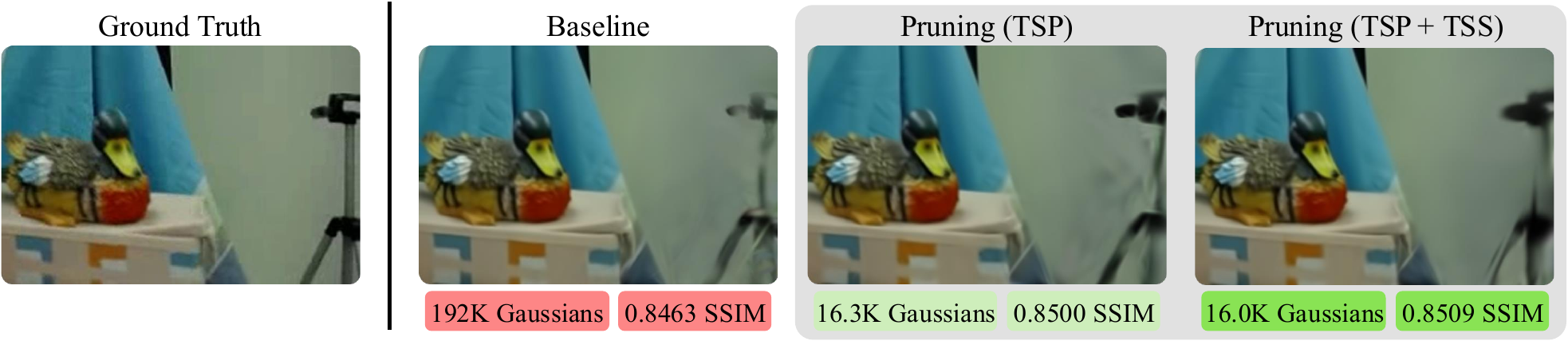}
  \vspace{-5mm}
\caption{\textbf{Comparison of our pruning methods on the real-world NeRF-DS~\cite{yan2023nerf} \emph{bell} scene.} 
Our proposed Temporal Sensitivity Pruning (TSP) and Temporal Sensitivity Sampling (TSS) methods achieve higher SSIM than the baseline DeformableGS~\cite{yang2023deformable3dgs} model while using $11\times$ fewer Gaussians. 
The left regions of the renderings appear visually identical, while the right regions show that combining TSP with TSS significantly reduces temporal flicker and floating artifacts compared to both standard pruning and the unpruned baseline.}
\vspace{-3mm}
  \label{fig:pruning_comparison}
\end{figure*}

\subsection{Temporal Sensitivity Pruning}
\label{sec:method:pruning}

Recent studies have shown that 3D Gaussian Splatting models are heavily over-parameterized -- comparable visual fidelity can often be achieved with far fewer primitives~\cite{HansonTuPUP3DGS}.
Because the computational cost of deformable 3DGS scales linearly with the number of Gaussians, reducing redundant primitives is a direct path to accelerating both training and rendering.
Building on the gradient-based pruning method of Speedy-Splat~\cite{HansonSpeedy}, which achieves state-of-the-art efficiency in static 3DGS, we extend this concept to dynamic reconstruction.
Our formulation introduces a \textbf{temporal sensitivity pruning score} that quantifies each Gaussian’s cumulative influence on reconstruction across both space and time.

For Gaussian $\Set{G}_i$, we compute the second-order sensitivity of the $L_2$ reconstruction loss to perturbations of its projected contribution $g_i$ across all training views and timesteps:
\begin{equation}
\nabla_{g_i}^2 L_2 =
\sum_{\phi, t \in \Set{P}_{gt}}
\left(
\left( \nabla_{g_i} I_{\Set{G}_t}(\phi) \right)^2
+ (I_{\Set{G}_t}(\phi) - I_{gt})\nabla_{g_i}^2 I_{\Set{G}_t}(\phi)
\right).
\label{eq:l2_d2}
\end{equation}
As training converges and the residual term diminishes, this simplifies to our temporal sensitivity pruning score:
\begin{equation}
\tilde{U}_{\Set{G}_i} \approx
\nabla_{g_i}^2 L_2
\approx
\sum_{\phi, t \in \Set{P}_{gt}}
\left( \nabla_{g_i} I_{\Set{G}_t}(\phi) \right)^2.
\label{eq:hessian}
\end{equation}
Here, $\Set{P}_{gt}$ is the set of training poses and timestamps $(\phi, t)$, and $I_{\Set{G}_t}(\phi)$ is the rendered image for the deformed Gaussians $\Set{G}_t$ at time $t$ and pose $\phi$.
$g_i$ is the 2D projection of Gaussian $\Set{G}_i$ as defined in Equation~\ref{eq:2d_gaussian_scalar}, and $\nabla_{g_i} I_{\Set{G}_t}(\phi)$ is the image-space gradient readily available from the renderer’s backward pass.
Since deformation parameters $(\Delta\mu_t, \Delta r_t, \Delta s_t)$ vary across timesteps, these gradients inherently encode temporal motion coupling, making $\tilde{U}_{\Set{G}_i}$ sensitive not only to static appearance but also to dynamic contribution. 

Our \textbf{Temporal Sensitivity Pruning (TSP)} approach periodically removes Gaussians with low temporal sensitivity pruning scores, extending sensitivity-based pruning to dynamic 3DGS. By aggregating second-order sensitivities over both spatial and temporal domains, TSP identifies motion-dependent redundancies and eliminates them in a principled, temporally aware manner. This reduces training and inference cost while preserving visual fidelity by concentrating computation on the most informative Gaussians.

\subsubsection{Temporal Sensitivity Sampling}
\label{sec:method:tss}

While TSP effectively identifies redundant primitives, it evaluates gradients only at the observed training timesteps.
This limitation mirrors prior pruning methods for \emph{static} 3DGS, which compute sensitivities from fixed viewpoints and single frames.
In dynamic scenes, however, a Gaussian’s contribution can vary over time, and gradient estimates confined to training frames may overlook temporally unstable primitives.
These unstable Gaussians -- often manifesting as \textit{floaters} -- exhibit weak gradients on observed views but introduce visible artifacts at unseen timestamps.

Previous gradient-based analyses~\cite{HansonTuPUP3DGS,Jiang2023FisherRF,11094666} have shown that gradient magnitudes can implicitly capture Gaussian sensitivity or stability.
We extend this concept to dynamic 3DGS by interpreting low sensitivities as indicators of temporal inconsistency -- Gaussians that appear stable under observed motion but drift when extrapolated to unseen deformation.
To expose such instability, we introduce \textbf{Temporal Sensitivity Sampling (TSS)}, a mechanism that probes the temporal dimension during sensitivity estimation.

TSS injects a linearly annealed Gaussian perturbation into the timestamp input of the deformation function, enabling the model to evaluate sensitivities over nearby motion states:
\begin{equation}
\begin{gathered}
(\mu + \Delta\mu, r + \Delta r, s + \Delta s)
= \mathcal{D}(\mu, r, s, t + \mathcal{X}(i)), \\
\mathcal{X}(i)
= \mathcal{N}(0,1)\!\cdot\!\beta\!\cdot\!\Delta t\!\cdot\!(1-i/\tau)
\end{gathered}
\label{eq:perturbation}
\end{equation}
where $\mathcal{D}$ is the deformation function, $\mathcal{N}(0,1)$ is standard normal noise, $\beta{=}0.1$ controls perturbation magnitude, $\Delta t$ is the frame interval, and $\tau{=}20{,}000$ is the annealing period. In our TSS approach, we extend our TSP method by evaluating sensitivities at perturbed motion states defined by Equation~\ref{eq:perturbation}. Since Equation~\ref{eq:hessian} requires no ground-truth supervision, TSS can compute these sensitivities in a fully self-supervised manner by sampling any temporally jittered motion state. Appendix~\ref{sec:appendix:tss} ablates our choice of $(\beta{=}0.1,\tau{=}20{,}000)$.

This temporal perturbation reveals Gaussians that appear well-behaved on training frames but behave inconsistently under small motion shifts -- a hallmark of floaters.
Such primitives receive lower sensitivity scores and are pruned, improving spatial compactness and temporal coherence.
Stronger perturbations early in training encourage temporal exploration and the removal of floaters during densification, while annealing the noise to zero in later iterations focuses optimization on precise reconstruction of the observed frames.
This schedule balances temporal robustness with reconstruction accuracy, yielding stable and deterministic evaluations.

Empirically, TSS improves pruning stability, suppresses floating artifacts, and enhances temporal smoothness.
When combined with TSP, it produces sharper, more temporally consistent reconstructions than standard gradient-based pruning approaches, as shown in Figure~\ref{fig:pruning_comparison}.
The impact of TSP and TSS is further analyzed in Section~\ref{sec:experiments:speede3dgs}, demonstrating their complementary effects on efficiency and fidelity.

\begin{figure*}
    \centering
    \includegraphics[width=\linewidth]{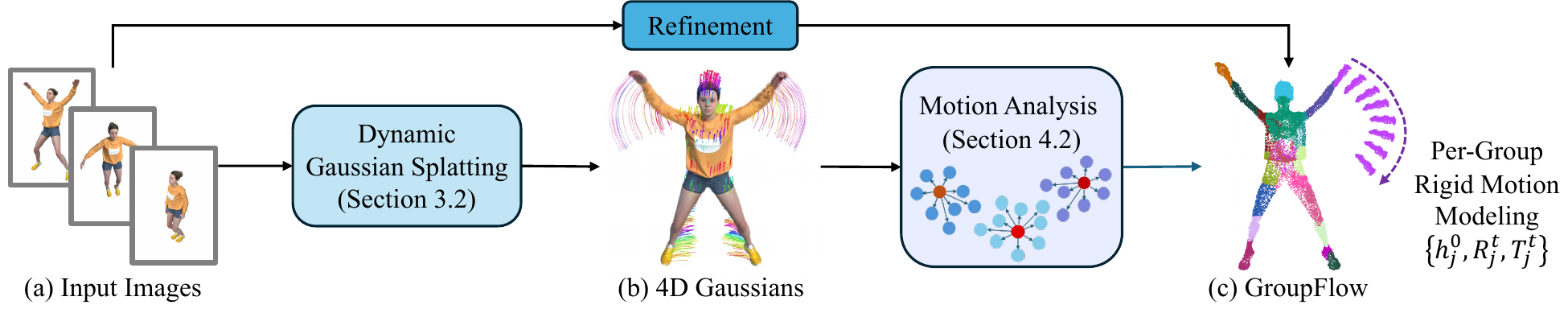}
    \vspace{-6mm}
    \caption{\textbf{Overview of our GroupFlow method.} Given a dynamic Gaussian Splatting model $\mathcal{G}$, we identify a subset of Gaussians as control points and assign each Gaussian to the control point $h_j$ with the most similar motion trajectory. The motion of each group is then estimated via a rigid transformation $[R_j^{t}|T_j^{t}]$ at each timestep, reducing inference from per-Gaussian to per-group deformation.
    }
    \label{fig:groupflow}
\vspace{-3mm}
\end{figure*}

\subsection{GroupFlow}
\label{sec:groupflow}

While pruning substantially reduces computation, dynamic Gaussian Splatting still incurs high inference cost because the deformation network must predict motion for every Gaussian.
A more efficient strategy is to distill learned neural motion into SE(3) transformations that encode rigid motion.
However, assigning a distinct SE(3) transformation to each Gaussian would be prohibitively expensive in both memory and parameter count, making this approach impractical for large or complex scenes.

To overcome this limitation, we introduce \textbf{GroupFlow}, illustrated in Figure~\ref{fig:groupflow}, which clusters Gaussians with similar motion trajectories and assigns each cluster a shared SE(3) transformation. Many dynamic objects in real scenes exhibit locally rigid motion, making groupwise modeling both natural and interpretable. By representing coherent motion with a single transformation rather than individual per-Gaussian predictions, GroupFlow reduces the number of motion trajectories from $N$ (one per Gaussian) to $J$ (one per group), balancing reconstruction fidelity and computational efficiency. This neural-to-rigid distillation yields a compact motion representation that accelerates rendering while maintaining temporal coherence and high visual quality.

\subsubsection{Flow Grouping via Motion Analysis}
\label{sec:flow_grouping}

To model coherent motion compactly, GroupFlow analyzes the trajectories of deformable Gaussians to identify clusters with similar motion. 
Starting from a dense deformable 3D Gaussian model $\mathcal{G}$, the motion $\mathcal{M}_i$ of each Gaussian $\mathcal{G}_i$ is represented as a sequence of means $\mu_i^t\in\mathbb{R}^3$ and rotation quaternions $r_i^t\in\mathbb{S}^3$ across $F$ timesteps:
\begin{equation}
    \mathcal{M} = \{\mathcal{M}_i\}_{i=1}^{N}, \quad
    \mathcal{M}_i = \{\mu_i^t, r_i^t\}_{t=0}^{F-1},
\end{equation}
where $N=|\mathcal{G}|$ is the total number of Gaussians. 
This unified 4D Gaussian representation enables similarity comparisons across trajectories.

We designate the first timestep $t=0$ as the canonical frame to ensure temporal consistency. To form motion-based groups, we select $J$ Gaussian means $h_j^t\in\mathbb{R}^3$ as control points via farthest point sampling on the set of means at $t=0$. All Gaussians $\mathcal{G}_i$ are then assigned to the control point $h_j$ with the most similar trajectory obtained via:
\begin{gather}
    \mathop{\arg\min}\limits_{j \in \{1, \dots, J\}} S_{i,j}, \\
    S_{i,j} = \lambda_r \mathrm{std}_{t}(\| \mu_i^t - h_j^t \|) + (1-\lambda_r) \mathrm{mean}_{t}(\| \mu_i^t - h_j^t \|), \nonumber
\end{gather}
where $S_{i,j}$ is the trajectory similarity score between $\mu_i$ and control point $h_j$, $\mathrm{std}_{t}(\cdot)$ and $\mathrm{mean}_{t}(\cdot)$ denote the standard deviation and mean of time-varying residual $\mu_i^t - h_j^t \in \mathbb{R}^3 $, and $\lambda_r=0.5$ is an empirically selected weighting ratio. All means assigned to the control point $h_j$ form a partitioning group $\mathcal{M}^j$ on $\mathcal{M}$.

Next, we estimate a time-varying rigid transformation for each group to capture its motion across time. For each group $j\in\{1,\dots,J\}$ and time $t>0$, we estimate an SE(3) rigid transformation $[R_j^{t}|T_j^{t}]$ that maps this group from the canonical frame at time $t=0$ to time $t$ using Umeyama alignment~\cite{sorkine2007arap, umeyama1991align}. Specifically, we randomly sample a subset of means $\mathcal{M}_{samp}^j$ ($|\mathcal{M}_{samp}^j| = \min\{N_j, N_{\max}\}$) from each group $\mathcal{M}^j$ and estimate their rigid transformation via:
\begin{equation}
\mathop{\arg\min}\limits_{R_j^{t}, T_j^{t}}
\sum_{\mu_i \in \mathcal{M}_{samp}^j}
\|   \mu_i^t - (
    R_j^{t} (\mu_i^0-h_j^0) + h_j^0 + T_j^{t}
    )\|^2,  
\end{equation}
where $\mu_i^t$ is the mean at time $t$, $\mu_i^0$ is the mean in the canonical frame at time $t=0$, $N_j=|\mathcal{M}^j|$ is the number of Gaussians in group $\mathcal{M}^j$, and $N_{\max} = 100$ is an empirically selected threshold representing the maximum number of means sampled per group. These group-wise motion parameters $\{h_j^0, R_j^{t}, T_j^{t}\}$ represent the shared flow of the mean $\mu_i$ -- and thus Gaussian $\mathcal{G}_i$ -- within group $\mathcal{M}^j$ over time.

\subsubsection{Flow Group Training and Inference}
To predict the motion parameters $\{\mu_i, r_i\} \in \mathcal{M}^j$ at timestep $t$, we apply an SE(3) rigid transformation $[R_j^{t}|T_j^{t}]$ to the canonical $\mu_i^0$, relative to its assigned control point $h_j^0$:
\begin{equation}
    \mu_i^{t} = R_j^{t} (\mu_i^0 - h_j^0) + h_j^0 + T_j^{t}.
\end{equation}
We also apply group-wise rotation $R_j^{t}$ to per-Gaussian rotation at canonical state $r_i^0$:
\begin{equation}
    r_i^t = \operatorname{quat}\!\big(R_j^t \operatorname{mat}(r_i^0)\big),
    \label{eq:per-gaussian-rotation}
\end{equation}
where $\operatorname{mat}(\cdot)$ converts quaternions to rotation matrices and $\operatorname{quat}(\cdot)$ performs the inverse. We set the shared flow $\{h_j^0, R_j^{t}, T_j^{t}\}$ of each group $\mathcal{M}^j$ as learnable parameters.

Our \textbf{GroupFlow} approach reduces the number of transformations predicted per timestep from $N$ (one per Gaussian) to $J$ (one per group) by enabling all Gaussians within a group to share the same motion parameters. 
In Appendix~\ref{sec:app:groups}, we ablate $J$ and select $J=2048$ groups to balance visual fidelity with total model size.

By distilling neural motion into grouped rigid transformations, GroupFlow significantly reduces deformation inference cost while maintaining high-quality motion reconstruction.
Beyond efficiency, GroupFlow also provides a regularizing effect on motion learning.
As discussed in Section~\ref{sec:experiments:monodygaubench}, it can stabilize reconstruction in scenes with noisy or inconsistent motion and even improve visual fidelity over the baseline despite using fewer motion parameters.
These results suggest that grouping similar motion trajectories not only compresses the deformation field but also enhances robustness to imperfect motion supervision.

\begin{table*}
\caption{\textbf{Results on the seven scenes in the real-world NeRF-DS dataset~\cite{yan2023nerf} with our SpeeDe3DGS framework.} \emph{TSP}, \emph{TSS}, and \emph{GF} denote Temporal Sensitivity Pruning, Sampling, and GroupFlow, respectively. \emph{Size} measures the combined deformation network and point cloud storage. Each experiment is run three times and averaged to reduce training variance. The \metrictablebest{best} and \metrictablesecond{second-best} results are highlighted. FPS and Train Time are measured on RTX~3090 and RTX~A5000 GPUs, respectively. Appendix~\ref{sec:appendix:scenes} reports per-scene results.}
\vspace{-1mm}
\centering
\resizebox{\linewidth}{!}{
\begin{tabular}{ccc|ccccccc}
\toprule
TSP & TSS & GF & PSNR~$\uparrow$ & SSIM~$\uparrow$ & LPIPS~$\downarrow$ & FPS~$\uparrow$ & Size (MB)~$\downarrow$ & \# Gaussians~$\downarrow$ & Train Time (s)~$\downarrow$ \\
\midrule
\multicolumn{3}{c|}{DeformableGS~\cite{yang2023deformable3dgs}} & \metrictablesecond{23.80} & 0.8503 & \metrictablebest{0.1781} & 54.37 (1.00$\times$) & 33.21 (1.00$\times$) & 132.22K (1.00$\times$) & 1523.83 (1.00$\times$) \\
    
 \checkmark &   &   & 23.78 & \metrictablesecond{0.8507} & 0.1863 & 346.96 (6.38x) & \metrictablebest{4.52 (7.35$\times$)} & \metrictablebest{10.90K (12.13$\times$)} & \metrictablesecond{741.66 (2.05$\times$)} \\
 
 \checkmark & \checkmark &   & \metrictablebest{23.81} & \metrictablebest{0.8515} & \metrictablesecond{0.1853} & 345.24 (6.35x) & \metrictablesecond{4.55 (7.29$\times$)} & \metrictablesecond{11.06K (11.95$\times$)} & 750.69 (2.03x) \\
 
    &   & \checkmark &  23.54 & 0.8433 & 0.1892 & \metrictablesecond{406.21 (8.58$\times$)} & 51.00 (0.65$\times$) & 132.32K (1.00$\times$) & 826.75 (1.84$\times$) \\
    
 \checkmark & \checkmark & \checkmark &  23.66 & 0.8487 & 0.1901 & \metrictablebest{505.60 (10.68$\times$)}   & 21.40 (1.55$\times$) & 11.10K (11.91$\times$) & \metrictablebest{625.48 (2.44$\times$)} \\
\bottomrule
\end{tabular}
}
\vspace{-3mm}
\label{tab:nerf_ds}
\end{table*}

\subsection{Speedy Deformable 3D Gaussian Splatting}
\label{sec:speede3dgs}

We integrate our methods into DeformableGS~\cite{yang2023deformable3dgs} to form \textbf{Speedy Deformable 3D Gaussian Splatting (SpeeDe3DGS)}, an efficient pipeline for dynamic scene reconstruction.
During training, we apply \textbf{Temporal Sensitivity Pruning (TSP)} and \textbf{Temporal Sensitivity Sampling (TSS)} to progressively adapt model capacity to scene complexity.
This two-phase strategy -- soft pruning during densification and hard pruning afterward -- removes redundant Gaussians and reduces deformation overhead, enabling real-time rendering while preserving the visual fidelity of neural motion models.
After densification, we apply \textbf{GroupFlow} to further accelerate inference by distilling the learned deformation field into grouped SE(3) transformations.
These shared groupwise motions replace per-Gaussian neural predictions, yielding rendering speeds that surpass prior dynamic Gaussian Splatting methods while maintaining high reconstruction quality and compact model size.
We ablate the components of SpeeDe3DGS and compare them with other dynamic Gaussian Splatting approaches in Section~\ref{sec:experiments}.
\section{Experiments}
\label{sec:experiments}

We evaluate our approach in two complementary settings:
(i) our \textbf{Speedy Deformable 3D Gaussian Splatting (SpeeDe3DGS) framework}, which extends DeformableGS~\cite{yang2023deformable3dgs} for detailed ablation and analysis, and
(ii) the \textbf{Monocular Dynamic Gaussian Splatting Benchmark (MonoDyGauBench)}~\cite{liang2025monocular}, which standardizes comparisons across motion representations.
Both settings integrate the complete SpeeDe3DGS pipeline -- Temporal Sensitivity Pruning (TSP), Temporal Sensitivity Sampling (TSS), and GroupFlow -- into dynamic Gaussian Splatting.

For consistent runtime comparison, all FPS metrics are measured on an NVIDIA RTX~3090 GPU.
Unless otherwise specified, experiments follow each framework’s default hyperparameters and train for a total of 30{,}000 iterations.
Starting at iteration 6{,}000, we apply TSP every 3{,}000 iterations, soft pruning 60\% of Gaussians at a time during densification and hard pruning 30\% after densification.
The impact of different soft and hard pruning percentages is analyzed in Appendix~\ref{sec:appendix:percentages}.
Temporal perturbations for TSS are annealed using $\beta{=}0.1$ and $\tau{=}20{,}000$, and GroupFlow is initialized after densification at iteration~15{,}000 with $J{=}2048$ motion groups.
The influence of group count on performance and model size is examined in Appendix~\ref{sec:app:groups}.
This configuration enables detailed ablations within SpeeDe3DGS and ensures fair cross-method comparisons under MonoDyGauBench.

\subsection{SpeeDe3DGS Results}
\label{sec:experiments:speede3dgs}

We first evaluate the components of SpeeDe3DGS on the real-world NeRF-DS dataset~\cite{yan2023nerf} in Table~\ref{tab:nerf_ds}.
Applying \textbf{TSP} alone accelerates rendering by $6.38\times$, reduces the number of Gaussians by $12.13\times$, and shortens training time by $2.05\times$ on average, while maintaining the image quality of DeformableGS~\cite{yang2023deformable3dgs}.
Adding \textbf{TSS} further improves PSNR and SSIM beyond the baseline, as temporal perturbations act as a regularizer that suppresses floaters and stabilizes motion.
Qualitative comparisons in Figure~\ref{fig:pruning_comparison} show that TSP and TSS together effectively remove floaters and improve temporal coherence without sacrificing sharpness.

Using \textbf{GroupFlow} without pruning yields an $8.58\times$ rendering speedup with a modest drop in image quality, reflecting the trade-off of representing motion through grouped rigid transformations rather than full neural deformation.
When combined with pruning (\textbf{TSP + TSS + GroupFlow}), rendering accelerates by $10.68\times$ and training time shortens by $2.44\times$.
Notably, the pruned representation improves GroupFlow’s motion grouping, achieving higher PSNR and SSIM than GroupFlow alone.
Even with additional GroupFlow parameters, the full SpeeDe3DGS configuration remains $1.55\times$ more compact than the baseline model.

Results on the synthetic D-NeRF~\cite{pumarola2020dnerf} and real-world HyperNeRF~\cite{park2021hypernerf} datasets are provided in Appendix~\ref{sec:app:additional_speede3dgs}.
On HyperNeRF, TSP + TSS and GroupFlow individually achieve $9.37\times$ and $15.66\times$ rendering speedups, respectively. When used together, they reach $29.21\times$ faster rendering, $12.18\times$ Gaussian reduction, and $3.74\times$ shorter training time, demonstrating the synergy of our methods.
Training times for NeRF-DS and D-NeRF are measured on RTX~A5000 GPUs (24~GB).
We use RTX~A6000 GPUs (48~GB) to train HyperNeRF due to the baseline memory requirements; however, pruning enables efficient training on 24~GB of VRAM. Per-scene metrics for all datasets are provided in Appendix~\ref{sec:appendix:scenes}.

\begin{table*}
\centering
\caption{\textbf{Results on Monocular Dynamic Gaussian Splatting Benchmark (MonoDyGauBench)~\cite{liang2025monocular}.} Quantitative results averaged across five datasets and 50 scenes for all methods in Section~\ref{sec:background:deformable}. We \textbf{\emph{cumulatively}} apply our SpeeDe3DGS methods to the DeformableGS~\cite{yang2023deformable3dgs} and 4DGS~\cite{Wu2024CVPR4dgs} baselines, keeping the original neural variants with \metrictableslow{low FPS} for reference, but excluding them from comparisons to focus on real-time methods. Pruning is performed using TSP and TSS. Each experiment is repeated three times and averaged. The \metrictablebest{best} and \metrictablesecond{second-best} results are highlighted; improvements over corresponding baselines are \textbf{bolded}. FPS and baseline Train Time are measured on an RTX~3090 GPU, while our Train Time* is measured on an RTX~A5000 (both 24~GB). Per-dataset results are reported in Appendix~\ref{sec:app:monodygaubench}.}
\vspace{-1mm}
\label{tab:all_methods_scenes_metrics}
\begin{tabular}{l|cccccc}
\toprule
Method & PSNR~$\uparrow$ & SSIM~$\uparrow$ & MS-SSIM~$\uparrow$ & LPIPS~$\downarrow$ & FPS~$\uparrow$ & Train Time (s)~$\downarrow$ \\
\midrule
EffGS~\cite{katsumata2023efficient} & 21.84 & 0.672 & 0.725 & 0.347 & 177.21 & 3757.81 \\
STG-decoder~\cite{li2023spacetime} & 21.81 & 0.678 & 0.742 & 0.352 & 109.42 & 5980.64 \\
STG~\cite{li2023spacetime} & 19.51 & 0.583 & 0.643 & 0.475 & 181.70 & 5359.56 \\
RTGS~\cite{yang2023ICLR4dgs} & 21.61 & 0.663 & 0.720 & 0.350 & 143.37 & 7352.52 \\
\midrule
4DGS~\cite{Wu2024CVPR4dgs} & \metrictableneural{23.55} & \metrictableneural{0.708} & \metrictableneural{0.765} & \metrictableneural{0.277} & \metrictableslow{62.99 (1.00$\times$)} & \metrictableneural{8628.89 (1.00$\times$)} \\
+ Pruning (Ours) & 22.44 & 0.689 & 0.737 & 0.334 & \textbf{179.64 (2.85$\times$)} & \textbf{4358.17* (1.47$\times$)}\\
+ GroupFlow (Ours) & 21.00 & 0.667 & 0.705 & 0.380 & \metrictablebest{\textbf{290.21 (4.61$\times$)}} & \textbf{4176.49* (2.07$\times$)} \\
\midrule
DeformableGS~\cite{yang2023deformable3dgs} & \metrictableneural{24.07} & \metrictableneural{0.694} & \metrictableneural{0.755} & \metrictableneural{0.283} & \metrictableslow{20.20 (1.00$\times$)} & \metrictableneural{6227.43 (1.00$\times$)} \\
+ Pruning (Ours) & \metrictablebest{23.86} & \metrictablesecond{0.694} & \metrictablesecond{0.749} & \metrictablebest{0.295} & \textbf{137.01 (6.78$\times$)} & \metrictablesecond{\textbf{2850.60* (2.18$\times$)}}\\
+ GroupFlow (Ours) & \metrictablesecond{23.52} & \metrictablebest{\textbf{0.709}} & \metrictablebest{\textbf{0.771}} & \metrictablesecond{0.313} & \metrictablesecond{\textbf{276.91 (13.71$\times$)}} & \metrictablebest{\textbf{2461.14* (2.53$\times$)}} \\
\bottomrule
\end{tabular}
\vspace{-4mm}
\label{tab:monodygaubench}
\end{table*}

\subsection{MonoDyGauBench Evaluation}
\label{sec:experiments:monodygaubench}

We evaluate our approach on the 50 scenes in \textbf{MonoDyGauBench}~\cite{liang2025monocular}, extending the high-fidelity yet computationally heavy DeformableGS~\cite{yang2023deformable3dgs} and 4DGS~\cite{Wu2024CVPR4dgs} baselines with our \textbf{Pruning (TSP + TSS)} and \textbf{GroupFlow} modules.
Table~\ref{tab:monodygaubench} summarizes aggregate results across all scenes, while Appendix~\ref{sec:app:monodygaubench} provides per-dataset breakdowns. Qualitative results are presented in Appendix~\ref{sec:app:mdgb_qualitative}.
Although MonoDyGauBench experiments are nominally conducted on an RTX~3090 GPU, several baseline scenes exceed its 24~GB memory limit -- a trend also noted in Section~\ref{sec:experiments:speede3dgs}.
Our pruning integration markedly reduces memory usage, enabling training within 24~GB on an RTX~A5000 GPU.

As shown in Table~\ref{tab:monodygaubench}, pruning makes \textbf{4DGS} on par with the fastest non-neural baselines while maintaining high image quality across metrics.
Adding GroupFlow on top of pruning achieves the highest rendering speed -- over 100~FPS faster than the fastest baseline -- but with some loss of image quality, suggesting that per-Gaussian inference in DeformableGS better preserves fidelity under grouped motion.

Integrating pruning into \textbf{DeformableGS} accelerates rendering by $6.78\times$, reaching frame rates comparable to RTGS~\cite{lin2024rtgs} while maintaining high visual fidelity across all datasets.
It surpasses all non-neural baselines and our 4DGS variants in overall image quality, reinforcing DeformableGS as a strong backbone for neural motion modeling.
Adding GroupFlow on top of pruning further increases rendering speed to $13.71\times$, making it nearly 100~FPS faster than all baseline methods and on-par with our 4DGS+Pruning+GroupFlow variant.
Training time is also the shortest, reduced by $2.53\times$.
Despite these speedups, reconstruction quality remains high: SSIM and MS-SSIM exceed the original DeformableGS baseline, suggesting a mild regularizing effect from GroupFlow.
As discussed in Appendix~\ref{sec:app:monodygaubench}, this effect is most pronounced in scenes with unstable pose estimation.
Overall, our SpeeDe3DGS strategies achieve an effective balance between speed, fidelity, and efficiency for dynamic Gaussian Splatting.
\section{Limitations}
\label{sec:limitations}

While SpeeDe3DGS achieves substantial acceleration with minimal fidelity loss, some trade-offs remain. Minor quality degradation can occur under extreme pruning ratios, though this can be mitigated by adjusting pruning hyperparameters to balance speed and fidelity. Similarly, GroupFlow preserves quality in most scenes through locally rigid motion, but highly deformable regions may lose fidelity when the number of motion groups is limited. We analyze these trade-offs through ablations of pruning percentages and number of groups in Appendices~\ref{sec:appendix:percentages}~and~\ref{sec:app:groups}. 
While our methods introduce mild regularizing effects that improve temporal stability, they do not incorporate explicit motion priors or learned dynamics. They are therefore complementary to prior-driven or motion-aware 3DGS frameworks and can accelerate such methods. In practice, TSP and TSS can be applied to any dynamic 3DGS framework, while GroupFlow extends to those with an explicit motion model. Appendix~\ref{sec:appendix:limitations} further discusses baseline and motion model limitations. For a broader analysis of dynamic Gaussian Splatting methods and their limitations, we refer readers to MonoDyGauBench~\cite{liang2025monocular}.
\section{Conclusion}
\label{sec:conclusion}

We introduced \textbf{Speedy Deformable 3D Gaussian Splatting (SpeeDe3DGS)}, a framework that bridges the efficiency–fidelity gap in dynamic 3D Gaussian Splatting by reducing redundant neural inference through pruning and motion distillation. 
Our \textbf{Temporal Sensitivity Pruning (TSP)} module removes redundant Gaussians via temporally aggregated sensitivity analysis, while \textbf{Temporal Sensitivity Sampling (TSS)} stabilizes this process by perturbing timestamps to suppress floaters and enhance temporal coherence. 
Complementing pruning, \textbf{GroupFlow} distills neural motion into shared rigid SE(3) transformations, enabling efficient groupwise deformation and faster convergence. 
Across the 50 dynamic scenes in MonoDyGauBench~\cite{liang2025monocular}, pruning accelerates DeformableGS by $6.78\times$ in rendering speed with minimal fidelity loss and $10\times$ fewer primitives, and adding GroupFlow further improves performance to $13.71\times$ faster rendering and $2.53\times$ shorter training while preserving superior image quality. 
Together, these modules demonstrate that temporally aware pruning and groupwise motion distillation can deliver neural-field fidelity at real-time speeds.
\section*{Acknowledgments}
\label{sec:acknowledgments}
This research is based upon work supported by the Office of the Director of National Intelligence (ODNI), Intelligence Advanced Research Projects Activity (IARPA), via IARPA R\&D Contract No. 140D0423C0076. The views and conclusions contained herein are those of the authors and should not be interpreted as necessarily representing the official policies or endorsements, either expressed or implied, of the ODNI, IARPA, or the U.S. Government. The U.S. Government is authorized to reproduce and distribute reprints for Governmental purposes notwithstanding any copyright annotation thereon. Commercial support was provided by the Amazon Research Awards program and Open Philanthropy. Further support was provided by DARPA TIAMAT and the NSF TRAILS Institute (2229885).

{
    \small
    \bibliographystyle{ieeenat_fullname}
    \bibliography{main}
}

\appendix
\section{Appendix}

\subsection{Per-Dataset Metrics for MonoDyGauBench}
\label{sec:app:monodygaubench}

We analyze results on Monocular Dynamic Gaussian Splatting Benchmark (MonoDyGauBench)~\cite{liang2025monocular}, which evaluates dynamic 3DGS variants across five datasets. For consistency, we directly use the baseline tables from the manuscript. Both \textbf{4DGS} and \textbf{DeformableGS} -- the neural motion-field baselines -- are extended with our pruning (TSP+TSS) and GroupFlow modules. Performance differences with our SpeeDe3DGS framework arise because MonoDyGauBench standardizes all methods within a unified training wrapper, integrating every motion representation and our components under common hyperparameters. Although MonoDyGauBench experiments are nominally conducted on an RTX~3090 GPU, baseline scenes in the HyperNeRF, Nerfies, and iPhone datasets exceed its 24~GB memory limit. Our pruning integration markedly reduces memory usage, enabling training within 24~GB on an RTX~A5000 GPU. Qualitative results are presented in Appendix~\ref{sec:app:mdgb_qualitative}.

On \textbf{D-NeRF}~\cite{pumarola2020dnerf}, Table~\ref{tab:mdgb:dnerf} shows that pruning and GroupFlow preserve the high neural fidelity of both baselines, surpassing all non-neural methods by a wide margin. Interestingly, adding GroupFlow slightly decreases rendering and training speed for DeformableGS -- a behavior unique to D-NeRF within MonoDyGauBench and not observed in our SpeeDe3DGS framework results in Appendix~\ref{sec:app:additional_speede3dgs}.

On \textbf{HyperNeRF}~\cite{park2021hypernerf}, Table~\ref{tab:mdgb:hypernerf} shows that adding pruning to DeformableGS improves PSNR and SSIM beyond the baseline with an $8.00\times$ rendering speedup, while adding GroupFlow further enhances SSIM and MS-SSIM, culminating in a $23.84\times$ speedup. Efficiency improvements with our SpeeDe3DGS framework on HyperNeRF are reported in Appendix~\ref{sec:app:additional_speede3dgs}.

On \textbf{NeRF-DS}~\cite{yan2023nerf}, Table~\ref{tab:mdgb:nerfds} shows that adding TSP and TSS yields comparable or improved image quality over the DeformableGS baseline, while adding GroupFlow provides a $10.49\times$ rendering acceleration with minimal fidelity loss. These trends are consistent with our findings in Section~\ref{sec:experiments:speede3dgs}. The 4DGS baseline performs comparatively poorly, consistent with MonoDyGauBench’s broader analysis of motion–representation trade-offs.

On \textbf{Nerfies}~\cite{park2021nerfies}, Table~\ref{tab:mdgb:nerfies} shows that pruning improves all quality metrics for DeformableGS, demonstrating the stabilizing effects of TSP+TSS. Adding GroupFlow further increases efficiency, achieving a $19.50\times$ rendering speedup and the highest FPS in the table, while maintaining higher image quality than all baselines.

On \textbf{iPhone}~\cite{gao2022dynamic}, Table~\ref{tab:mdgb:iphone} shows that adding GroupFlow significantly improves PSNR, SSIM, and MS-SSIM while achieving a $26.34\times$ rendering speedup. By enforcing locally consistent motion through shared rigid SE(3) transformations, GroupFlow acts as a strong structural regularizer that enhances temporal coherence and mitigates drift under noisy or unstable camera trajectories.

\begin{table}[t]
\centering
\caption{\textbf{Results for the eight scenes in the synthetic D-NeRF dataset~\cite{pumarola2020dnerf} with MonoDyGauBench~\cite{liang2025monocular}.} Quantitative results for all methods in Section~\ref{sec:background:deformable}. We \textbf{\emph{cumulatively}} apply our SpeeDe3DGS methods to the DeformableGS~\cite{yang2023deformable3dgs} and 4DGS~\cite{Wu2024CVPR4dgs} baselines, keeping the original neural variants with \metrictableslow{low FPS} for reference, but excluding them from comparisons to focus on real-time methods. Pruning is performed using TSP and TSS. Each experiment is repeated three times and averaged. The \metrictablebest{best} and \metrictablesecond{second-best} results are highlighted; improvements over corresponding baselines are \textbf{bolded}. FPS and baseline Train Time are measured on an RTX~3090 GPU, while our Train Time* is measured on an RTX~A5000 (both 24~GB). Results with our SpeeDe3DGS framework are reported in Table~\ref{tab:d_nerf}.}
\vspace{-2mm}
\resizebox{\columnwidth}{!}{
\begin{tabular}{l|cccccc}
\toprule
Method & PSNR~$\uparrow$ & SSIM~$\uparrow$ & MS-SSIM~$\uparrow$ & LPIPS~$\downarrow$ & FPS~$\uparrow$ & Train Time (s)~$\downarrow$ \\
\midrule
EffGS~\cite{katsumata2023efficient} & 30.52 & 0.96 & 0.98 & 0.04 & \metrictablesecond{289.47} & \metrictablebest{1042.04} \\
STG-decoder~\cite{li2023spacetime} & 25.89 & 0.91 & 0.90 & 0.17 & 160.32 & 3462.29 \\
STG~\cite{li2023spacetime} & 17.09 & 0.88 & 0.66 & 0.29 & 208.98 & 4889.50 \\
RTGS~\cite{yang2023ICLR4dgs} & 28.78 & 0.96 & 0.96 & 0.05 & 192.37 & 1519.60 \\
\midrule
4DGS~\cite{Wu2024CVPR4dgs} & \metrictableneural{33.27} & \metrictableneural{0.98} & \metrictableneural{0.99} & \metrictableneural{0.02} & \metrictableslow{134.13 (1.00$\times$)} & \metrictableneural{1781.46 (1.00$\times$)} \\
+ Pruning (Ours) & 32.99 & 0.98 & 0.99 & 0.02 & \textbf{238.98 (1.78$\times$)} & \textbf{1263.30* (1.41$\times$)} \\
+ GroupFlow (Ours) & 31.70 & 0.97 & 0.99 & 0.03 & \textbf{264.61 (1.97$\times$)} & \textbf{1296.21* (1.37$\times$)}\\
\midrule
DeformableGS~\cite{yang2023deformable3dgs} & \metrictableneural{37.14} & \metrictableneural{0.99} & \metrictableneural{0.99} & \metrictableneural{0.01} & \metrictableslow{50.78 (1.00$\times$)} & \metrictableneural{2048.38 (1.00$\times$)} \\
+ Pruning (Ours) & \metrictablebest{36.25} & \metrictablebest{0.99} & \metrictablebest{0.99} & \metrictablebest{0.01} & \metrictablebest{\textbf{294.44 (5.80$\times$)}} & \textbf{1241.56* (1.65$\times$)} \\
+ GroupFlow (Ours) & \metrictablesecond{35.18} & \metrictablesecond{0.99} & \metrictablesecond{0.99} & \metrictablesecond{0.01} & \textbf{275.32 (5.42$\times$)} & \metrictablesecond{\textbf{1164.55* (1.76$\times$)}} \\
\bottomrule
\end{tabular}
}
\label{tab:mdgb:dnerf}
\vspace{-1mm}
\end{table}

\begin{table}[t]
\centering
\caption{\textbf{Results for the 17 scenes in the real-world HyperNeRF dataset~\cite{park2021hypernerf} with MonoDyGauBench~\cite{liang2025monocular}.} Results with our SpeeDe3DGS framework are reported in Table~\ref{tab:hypernerf}.}
\vspace{-2mm}
\resizebox{\columnwidth}{!}{
\begin{tabular}{l|cccccc}
\toprule
Method & PSNR~$\uparrow$ & SSIM~$\uparrow$ & MS-SSIM~$\uparrow$ & LPIPS~$\downarrow$ & FPS~$\uparrow$ & Train Time (s)~$\downarrow$ \\
\midrule
EffGS~\cite{katsumata2023efficient} & 22.24 & 0.70 & 0.79 & 0.37 & 138.79 & 4119.66 \\
STG-decoder~\cite{li2023spacetime} & 23.92 & 0.73 & 0.83 & 0.34 & 66.72 & 7423.41 \\
STG~\cite{li2023spacetime} & 22.92 & 0.70 & 0.78 & 0.41 & 183.21 & 5729.80 \\
RTGS~\cite{yang2023ICLR4dgs} & 22.99 & 0.71 & 0.79 & 0.35 & 104.64 & 11507.80 \\
\midrule
4DGS~\cite{Wu2024CVPR4dgs} & \metrictableneural{25.70} & \metrictableneural{0.79} & \metrictableneural{0.89} & \metrictableneural{0.23} & \metrictableslow{37.46 (1.00$\times$)} & \metrictableneural{10170.86 (1.00$\times$)} \\
+ Pruning (Ours) & \metrictablebest{25.47} & \metrictablebest{0.78} & \metrictablebest{0.88} & \metrictablebest{0.28} & \textbf{157.91 (4.22$\times$)} & \textbf{5623.29* (1.81$\times$)} \\
+ GroupFlow (Ours) & 23.98 & 0.75 & 0.85 & \metrictablesecond{0.31} & \metrictablebest{\textbf{269.57 (7.20$\times$)}} & \textbf{5871.93* (1.73$\times$)}\\
\midrule
DeformableGS~\cite{yang2023deformable3dgs} & \metrictableneural{24.58} & \metrictableneural{0.74} & \metrictableneural{0.83} & \metrictableneural{0.27} & \metrictableslow{10.91 (1.00$\times$)} & \metrictableneural{8855.46 (1.00$\times$)} \\
+ Pruning (Ours) & \metrictablesecond{\textbf{24.83}} & \textbf{0.75} & 0.82 & 0.31 & \textbf{87.27 (8.00$\times$)} & \metrictablesecond{\textbf{3668.02* (2.41$\times$)}} \\
+ GroupFlow (Ours) & 24.13 & \metrictablesecond{\textbf{0.75}} & \metrictablesecond{\textbf{0.85}} & 0.32 & \metrictablesecond{\textbf{260.05 (23.84$\times$)}} & \metrictablebest{\textbf{3361.70* (2.63$\times$)}}\\
\bottomrule
\end{tabular}
}
\label{tab:mdgb:hypernerf}
\vspace{-1mm}
\end{table}

\begin{table}[t]
\centering
\caption{\textbf{Results for the seven scenes in the real-world NeRF-DS dataset~\cite{yan2023nerf} with MonoDyGauBench~\cite{liang2025monocular}.} Results with our SpeeDe3DGS framework are reported in Table~\ref{tab:nerf_ds}.}
\vspace{-2mm}
\resizebox{\columnwidth}{!}{
\begin{tabular}{l|cccccc}
\toprule
Method & PSNR~$\uparrow$ & SSIM~$\uparrow$ & MS-SSIM~$\uparrow$ & LPIPS~$\downarrow$ & FPS~$\uparrow$ & Train Time (s)~$\downarrow$ \\
\midrule
EffGS~\cite{katsumata2023efficient} & 21.28 & 0.78 & 0.77 & 0.25 & 307.70 & 1597.90 \\
STG-decoder~\cite{li2023spacetime} & 21.73 & 0.80 & 0.81 & 0.20 & 212.24 & 2131.48 \\
STG~\cite{li2023spacetime} & 20.13 & 0.69 & 0.72 & 0.39 & 302.89 & 2214.62 \\
RTGS~\cite{yang2023ICLR4dgs} & 19.88 & 0.75 & 0.73 & 0.30 & 259.83 & 3116.21 \\
\midrule
4DGS~\cite{Wu2024CVPR4dgs} & \metrictableneural{20.08} & \metrictableneural{0.73} & \metrictableneural{0.71} & \metrictableneural{0.25} & \metrictableslow{100.22 (1.00$\times$)} & \metrictableneural{4075.43 (1.00$\times$)} \\
+ Pruning (Ours) & \textbf{20.27} & \textbf{0.74} & 0.73 & 0.24 & \textbf{255.48 (2.55$\times$)} & \textbf{2787.93* (1.46$\times$)}\\
+ GroupFlow (Ours) & 20.14 & 0.73 & 0.72 & 0.25 & \metrictablebest{\textbf{323.86 (3.23$\times$)}} & \textbf{2712.69* (1.50$\times$)} \\
\midrule
DeformableGS~\cite{yang2023deformable3dgs} & \metrictableneural{23.42} & \metrictableneural{0.84} & \metrictableneural{0.88} & \metrictableneural{0.16} & \metrictableslow{30.27 (1.00$\times$)} & \metrictableneural{2885.07 (1.00$\times$)} \\
+ Pruning (Ours) & \metrictablebest{\textbf{23.49}} & \metrictablebest{0.84} & \metrictablebest{\textbf{0.89}} & \metrictablebest{\textbf{0.15}} & \textbf{255.48 (8.44$\times$)} & \metrictablebest{\textbf{1488.77* (1.94$\times$)}} \\
+ GroupFlow (Ours) & \metrictablesecond{23.36} & \metrictablesecond{0.84} & \metrictablesecond{0.88} & \metrictablesecond{\textbf{0.15}} & \metrictablesecond{\textbf{317.47 (10.49$\times$)}} & \metrictablesecond{\textbf{1556.77* (1.85$\times$)}}\\
\bottomrule
\end{tabular}
}
\label{tab:mdgb:nerfds}
\vspace{-4mm}
\end{table}

\begin{table}[h]
\centering
\caption{\textbf{Results for the four scenes in the real-world Nerfies dataset~\cite{park2021nerfies} with MonoDyGauBench~\cite{liang2025monocular}.}}
\vspace{-2mm}
\resizebox{\columnwidth}{!}{
\begin{tabular}{l|cccccc}
\toprule
Method & PSNR~$\uparrow$ & SSIM~$\uparrow$ & MS-SSIM~$\uparrow$ & LPIPS~$\downarrow$ & FPS~$\uparrow$ & Train Time (s)~$\downarrow$ \\
\midrule
EffGS~\cite{katsumata2023efficient} & 20.13 & 0.43 & 0.61 & 0.65 & 159.17 & \metrictablesecond{3249.33} \\
STG-decoder~\cite{li2023spacetime} & 20.93 & \metrictablesecond{0.47} & 0.67 & 0.58 & 90.46 & 6050.42 \\
STG~\cite{li2023spacetime} & 20.55 & 0.46 & 0.66 & 0.62 & 142.30 & 5264.58 \\
RTGS~\cite{yang2023ICLR4dgs} & 20.06 & 0.42 & 0.62 & 0.61 & 153.38 & 9059.54 \\
\midrule
4DGS~\cite{Wu2024CVPR4dgs} & \metrictableneural{21.84} & \metrictableneural{0.50} & \metrictableneural{0.72} & \metrictableneural{0.47} & \metrictableslow{35.70 (1.00$\times$)} & \metrictableneural{9797.50 (1.00$\times$)} \\
+ Pruning (Ours) & \metrictablebest{21.79} & \metrictablebest{0.49} & \metrictablebest{0.71} & \metrictablesecond{0.55} & \textbf{143.74 (4.03$\times$)} & \textbf{5760.44* (1.70$\times$)} \\
+ GroupFlow (Ours) & 20.59 & 0.46 & 0.66 & 0.58 & \metrictablesecond{\textbf{280.14 (7.85$\times$)}} & \textbf{5776.24* ($\times$)} \\
\midrule
DeformableGS~\cite{yang2023deformable3dgs} & \metrictableneural{21.26} & \metrictableneural{0.43} & \metrictableneural{0.66} & \metrictableneural{0.54} & \metrictableslow{14.99 (1.00$\times$)} & \metrictableneural{6950.94 (1.00$\times$)} \\
+ Pruning (Ours) & \metrictablesecond{\textbf{21.76}} & \textbf{0.46} & \metrictablesecond{\textbf{0.69}} & \metrictablebest{\textbf{0.51}} & \textbf{85.21 (5.68$\times$)} & \textbf{3935.12* (1.77$\times$)} \\
+ GroupFlow (Ours) & 21.35 & \textbf{0.46} & 0.67 & 0.58 & \metrictablebest{\textbf{284.86 (19.00$\times$)}} & \metrictablebest{\textbf{3001.68* (2.32$\times$)}} \\
\bottomrule
\end{tabular}
}
\label{tab:mdgb:nerfies}
\vspace{-1mm}
\end{table}

\begin{table}[h]
\centering
\caption{\textbf{Results for the 14 scenes in the real-world iPhone dataset~\cite{gao2022dynamic} with MonoDyGauBench~\cite{liang2025monocular}.}}
\vspace{-2mm}
\resizebox{\columnwidth}{!}{
\begin{tabular}{l|cccccc}
\toprule
Method & PSNR~$\uparrow$ & SSIM~$\uparrow$ & MS-SSIM~$\uparrow$ & LPIPS~$\downarrow$ & FPS~$\uparrow$ & Train Time (s)~$\downarrow$ \\
\midrule
EffGS~\cite{katsumata2023efficient} & \metrictablesecond{16.82} & 0.47 & 0.50 & \metrictablesecond{0.47} & 99.64 & 6275.33 \\
STG-decoder~\cite{li2023spacetime} & \metrictablebest{16.85} & \metrictablesecond{0.47} & \metrictablesecond{0.52} & 0.49 & 86.18 & 7694.83 \\
STG~\cite{li2023spacetime} & 15.60 & 0.40 & 0.45 & 0.57 & 114.95 & 6629.64 \\
RTGS~\cite{yang2023ICLR4dgs} & 15.36 & 0.39 & 0.44 & 0.52 & 101.33 & 8484.05 \\
\midrule
4DGS~\cite{Wu2024CVPR4dgs} & \metrictableneural{15.13} & \metrictableneural{0.42} & \metrictableneural{0.43} & \metrictableneural{0.52} & \metrictableslow{42.53 (1.00$\times$)} & \metrictableneural{14205.48 (1.00$\times$)} \\
+ Pruning (Ours) & 14.00 & \textbf{0.44} & 0.43 & 0.56 & \textbf{144.45 (3.40$\times$)} & \textbf{4974.91* (2.86$\times$)} \\
+ GroupFlow (Ours) & 11.83 & 0.42 & 0.38 & 0.68 & \metrictablebest{\textbf{315.97 (7.43$\times$)}} & \textbf{4038.46* ($\times$)} \\
\midrule
DeformableGS~\cite{yang2023deformable3dgs} & \metrictableneural{16.56} & \metrictableneural{0.46} & \metrictableneural{0.48} & \metrictableneural{0.45} & \metrictableslow{10.47 (1.00$\times$)} & \metrictableneural{7278.28* (1.00$\times$)} \\
+ Pruning (Ours) & 16.40 & 0.45 & 0.47 & \metrictablebest{0.45} & \textbf{74.26 (7.09$\times$)} & \metrictablesecond{\textbf{3148.54* (2.31$\times$)}} \\
+ GroupFlow (Ours) & \textbf{16.82} & \metrictablebest{\textbf{0.51}} & \metrictablebest{\textbf{0.53}} & 0.48 & \metrictablesecond{\textbf{275.75 (26.34$\times$)}} & \metrictablebest{\textbf{2406.27* (3.02$\times$)}}\\
\bottomrule
\end{tabular}
}
\label{tab:mdgb:iphone}
\vspace{-4mm}
\end{table}

\subsection{Additional SpeeDe3DGS Results}
\label{sec:app:additional_speede3dgs}

We extend our analysis to the synthetic D-NeRF~\cite{pumarola2020dnerf} and real-world HyperNeRF~\cite{park2021hypernerf} datasets using the \textbf{DeformableGS}~\cite{yang2023deformable3dgs} codebase.
Rendering FPS is measured on an RTX~3090 GPU with 24~GB of VRAM for consistency with MonoDyGauBench~\cite{liang2025monocular}.

On \textbf{D-NeRF}~\cite{pumarola2020dnerf}, both pruning and GroupFlow individually provide around a $3\times$ increase in rendering speed and a $1.5\times$ reduction in training time, measured on an RTX~A5000 GPU (24~GB).
Pruning slightly reduces reconstruction quality more than GroupFlow, likely because D-NeRF’s synthetic scenes are compact, noise-free, and already sparsely parameterized, leaving limited redundancy to remove.
Nevertheless, pruning achieves a $13.79\times$ reduction in the number of Gaussians.
Since these scenes contain no pose noise or floaters, TSP and TSP+TSS perform nearly identically, showing that TSS adds stability without affecting clean synthetic settings.
GroupFlow performs particularly well due to the simple, rigid motions of D-NeRF scenes.
When combined, pruning and GroupFlow achieve a $4.11\times$ rendering speedup and a $1.93\times$ reduction in training time.

For \textbf{HyperNeRF}~\cite{park2021hypernerf}, we report results on the same eight real-world scenes used in DeformableGS -- \emph{espresso}, \emph{americano}, \emph{cookie}, \emph{chicken}, \emph{torchocolate}, \emph{lemon}, \emph{hand}, and \emph{printer}.
Following DeformableGS, image-quality metrics are not reported because most scenes lack evaluation splits; we provide full results on MonoDyGauBench in Table~\ref{tab:mdgb:hypernerf}.
Pruning alone yields a $9.37\times$ rendering speedup, an $11.88\times$ reduction in Gaussians, and a $2.67\times$ reduction in training time.
GroupFlow independently achieves a $15.66\times$ rendering speedup and a $2.10\times$ shorter training time.
Together, they reach $29.21\times$ faster rendering and $3.74\times$ shorter training, demonstrating strong synergy between our \acronym{} components.
Baseline HyperNeRF models require an RTX~A6000 GPU with 48~GB of VRAM for training, whereas our pruned configurations fit comfortably within 24~GB on an RTX~A5000.
All inference metrics, including FPS, are collected on an RTX~3090 GPU (24~GB), matching the MonoDyGauBench setup and underscoring that baseline models exceed 24~GB memory during training despite being evaluated on an RTX~3090 for inference.

\begin{table*}
\caption{\textbf{Results on the eight scenes in the synthetic D-NeRF dataset~\cite{pumarola2020dnerf} with our SpeeDe3DGS framework.} \emph{TSP}, \emph{TSS}, and \emph{GF} denote Temporal Sensitivity Pruning, Sampling, and GroupFlow, respectively. \emph{Size} measures the combined deformation network and point cloud storage. Each experiment is run three times and averaged to reduce training variance. The \metrictablebest{best} and \metrictablesecond{second-best} results are highlighted. FPS and Train Time are measured on RTX~3090 and RTX~A5000 GPUs, respectively. Per-scene results are provided in Appendix~\ref{sec:appendix:scenes}. Results with MonoDyGauBench~\cite{liang2025monocular} are reported in Table~\ref{tab:mdgb:dnerf}.}
\centering
\resizebox{\linewidth}{!}{
\begin{tabular}{ccc|ccccccc}
\toprule
TSP & TSS & GF & PSNR~$\uparrow$ & SSIM~$\uparrow$ & LPIPS~$\downarrow$ & FPS~$\uparrow$ & Size (MB)~$\downarrow$ & \# Gaussians~$\downarrow$ & Train Time (s)~$\downarrow$ \\
\midrule
\multicolumn{3}{c|}{DeformableGS~\cite{yang2023deformable3dgs}} & \metrictablebest{38.92} & \metrictablebest{0.9892} & \metrictablebest{0.0143} & 127.47 (1.00$\times$) & 16.88 (1.00$\times$) & 62.92K (1.00$\times$) & 940.52 (1.00$\times$) \\

 \checkmark &   &   & 36.19 & 0.9793 & 0.0349 & 422.48 (3.31$\times$) & \metrictablebest{3.08 (5.48$\times$)} & \metrictablesecond{4.56K (13.79$\times$)} & 524.52 (1.79$\times$) \\
 
 \checkmark & \checkmark &   & 36.19 & 0.9792 & 0.0350 & \metrictablesecond{423.89 (3.33$\times$)} & \metrictablesecond{3.08 (5.48$\times$)} & 4.57K (13.78$\times$) & \metrictablesecond{524.34 (1.79$\times$)} \\
 
    &   & \checkmark &  \metrictablesecond{36.85} & \metrictablesecond{0.9862} & \metrictablesecond{0.0172} & 374.32 (2.94$\times$) & 28.17 (0.60$\times$) & 62.77K (1.00$\times$) & 636.39 (1.48$\times$) \\
    
 \checkmark & \checkmark & \checkmark &  35.07 & 0.9771 & 0.0365 & \metrictablebest{524.19 (4.11$\times$)} & 13.96 (1.21$\times$) & \metrictablebest{4.56K (13.79$\times$)} & \metrictablebest{486.10 (1.93$\times$)} \\
\bottomrule
\end{tabular}
}
\label{tab:d_nerf}
\end{table*}

\begin{table*}
\caption{\textbf{Results on the eight real-world scenes from the HyperNeRF dataset~\cite{park2021hypernerf} in the DeformableGS~\cite{yang2023deformable3dgs} paper with our SpeeDe3DGS framework.} \emph{TSP}, \emph{TSS}, and \emph{GF} denote Temporal Sensitivity Pruning, Sampling, and GroupFlow, respectively. \emph{Size} measures the combined deformation network and point cloud storage. Each experiment is run three times and averaged to reduce training variance. The \metrictablebest{best} and \metrictablesecond{second-best} results are highlighted. FPS and Train Time are measured on RTX~3090 and RTX~A6000 GPUs, respectively. Per-scene results are reported in Appendix \ref{sec:appendix:scenes}. Results with MonoDyGauBench~\cite{liang2025monocular} are reported in Table~\ref{tab:mdgb:hypernerf}.}
\centering
\begin{tabular}{ccc|ccccc}
\toprule
TSP & TSS & GF & FPS~$\uparrow$ & Size (MB)~$\downarrow$ & \# Gaussians~$\downarrow$ & Train Time (s)~$\downarrow$ \\
\midrule
\multicolumn{3}{c|}{DeformableGS~\cite{yang2023deformable3dgs}}  &  14.48 (1.00$\times$) & 159.30  (1.00$\times$) & 665.35K (1.00$\times$) & 5248.91 (1.00$\times$)\\

\checkmark &  &   & 141.11 (9.75$\times$) & \metrictablebest{14.98 (10.64$\times$)} &  \metrictablesecond{55.12K (12.07$\times$)} & 1992.39 (2.63$\times$) \\

\checkmark & \checkmark &   & 135.70 (9.37$\times$) & \metrictablesecond{15.18 (10.49$\times$)} & 56.00K (11.88$\times$) & \metrictablesecond{1968.72 (2.67$\times$)}   \\
 
&  & \checkmark & \metrictablesecond{226.77 (15.66$\times$)} & 179.79 (0.89$\times$) & 666.31K (1.00$\times$) & 2505.44 (2.10$\times$) \\

\checkmark & \checkmark & \checkmark & \metrictablebest{422.88 (29.21$\times$)} & 31.62 (5.04$\times$) & \metrictablebest{54.63K (12.18$\times$)} & \metrictablebest{1401.85 (3.74$\times$)} \\
\bottomrule
\end{tabular}
\label{tab:hypernerf}
\end{table*}

\subsection{Ablation on Number of Groups}
\label{sec:app:groups}

We ablate the number of motion groups $J$ for GroupFlow on the NeRF-DS~\cite{yan2023nerf} and D-NeRF~\cite{pumarola2020dnerf} datasets using the DeformableGS~\cite{yang2023deformable3dgs} codebase.
Table~\ref{tab:nerf_ds_groups} reports quantitative results averaged over three independent runs to reduce training variance.
All FPS and training time measurements are collected on RTX~3090 and RTX~A5000 GPUs.

Increasing $J$ improves reconstruction quality by allowing finer motion grouping, though the number of learnable parameters grows linearly with $J$.
We select $J{=}2048$ as it provides the best balance between fidelity and compactness – the combined size of the pruned DeformableGS and GroupFlow model is smaller than that of the unpruned baseline and its 1.94~MB deformation network.
Smaller configurations, such as $J{=}128$, produce even lighter GroupFlow models than the deformation network but incur a minor drop in quality.
Conversely, $J{=}4096$ slightly degrades performance, suggesting that GroupFlow introduces a mild regularizing effect on SE(3) transformations, as nearby Gaussians naturally share coherent motion.
Rendering FPS remains nearly constant across all $J$ due to the efficiency of batched SE(3) transformations.
Notably, combining pruning with GroupFlow yields better results than GroupFlow alone across all $J$ values on the NeRF-DS dataset, likely because pruning suppresses noise that would otherwise hinder motion grouping.

\begin{table*}
\caption{\textbf{Ablation on group count $J$ on the NeRF-DS~\cite{yan2023nerf} and D-NeRF~\cite{pumarola2020dnerf} datasets with our SpeeDe3DGS framework.} $J{=}-$ indicates that GroupFlow is not used. Each experiment is repeated three times and averaged to reduce training variance. The \metrictablebest{best} and \metrictablesecond{second best} results are highlighted. FPS and Train Time are measured on RTX~3090 and RTX~A5000 GPUs, respectively. }
\centering
\resizebox{\linewidth}{!}{
\begin{tabular}{cc|ccccccc|ccccccc}
\toprule
& & \multicolumn{7}{c|}{{\small NeRF-DS~\cite{yan2023nerf}}} & \multicolumn{7}{c}{{\small D-NeRF~\cite{pumarola2020dnerf}}} \\
Prune & $J$ & PSNR~$\uparrow$ & SSIM~$\uparrow$ & LPIPS~$\downarrow$ & FPS~$\uparrow$ & Size (MB)~$\downarrow$ & \# Gaussians~$\downarrow$ & Train Time (s)~$\downarrow$ & PSNR~$\uparrow$ & SSIM~$\uparrow$ & LPIPS~$\downarrow$ & FPS~$\uparrow$ & Size (MB)~$\downarrow$ & \# Gaussians~$\downarrow$ & Train Time (s)~$\downarrow$\\
\midrule
& -- & \textit{23.80} & \textit{0.8503} & \textit{0.1781} & \textit{54.37} & \textit{33.21} & \textit{132.22K} & \textit{1523.83} & \textit{38.92} & \textit{0.9892} & \textit{0.0143} & \textit{127.47} & \textit{16.88} & \textit{62.92K} & \textit{940.52} \\
 & 128 & 22.67 & 0.8181 & 0.2322 & 415.53 & \metrictablebest{36.92} & 146.43K & 905.16 & 35.91 & 0.9844 & 0.0185 & \metrictablesecond{376.31} & \metrictablebest{16.22} & 63.13K & 651.65 \\
 & 256 & 23.04 & 0.8322 & 0.2103 & 407.14 & \metrictablesecond{37.61} & 144.46K & 916.39 & 36.28 & 0.9851 & 0.0179 & 373.42 & \metrictablesecond{16.97} & 62.92K & 643.34 \\
 & 512 & 22.86 & 0.8238 & 0.2232 & \metrictablebest{417.04} & 38.43 & 138.27K & 855.18 & 36.68 & 0.9860 & \metrictablebest{0.0170} & 375.79 & 18.66 & 63.27K & 640.09 \\
 & 1024 & \metrictablesecond{23.48} & \metrictablesecond{0.8421} & \metrictablesecond{0.1919} & 406.31 & 41.75 & \metrictablesecond{132.73K} & \metrictablesecond{827.25} & 36.75 & \metrictablesecond{0.9861} & \metrictablesecond{0.0170} & 373.09 & 21.81 & 63.02K & \metrictablesecond{639.20} \\
 & 2048 & \metrictablebest{23.54} & \metrictablebest{0.8433} & \metrictablebest{0.1892} & 406.21 & 51.00 & \metrictablebest{132.32K} & \metrictablebest{826.75} & \metrictablesecond{36.85} & \metrictablebest{0.9862} & 0.0172 & 374.32 & 28.17 & \metrictablesecond{62.77K} & \metrictablebest{636.39} \\
 & 4096 & 22.64 & 0.7995 & 0.2195 & \metrictablesecond{416.92} & 69.89 & 133.14K & 879.70 & \metrictablebest{36.96} & 0.9861 & 0.0177 & \metrictablebest{376.61} & 40.99 & \metrictablebest{62.71K} & 645.95 \\
\midrule
\checkmark & -- & \textit{23.81} & \textit{0.8515} & \textit{0.1853} & \textit{345.24}  & \textit{4.55} & \textit{11.06K}  & \textit{750.69} & \textit{36.19} & \textit{0.9792} & \textit{0.0350} & \textit{423.89} & \textit{3.08} & \textit{4.57K}  & \textit{524.34}  \\
\checkmark & 128 & 23.29 & 0.8402 & 0.2012 & \metrictablesecond{514.64} & \metrictablebest{3.86} & \metrictablesecond{11.01K} & 635.63 & 34.64 & 0.9767 & 0.0363 & 519.12 & \metrictablebest{1.92} & 4.56K & 496.64 \\
\checkmark & 256 & 23.41 & 0.8440 & 0.1958 & \metrictablebest{516.81} & \metrictablesecond{5.05} & 11.09K & 665.38 & 34.96 & 0.9773 & \metrictablesecond{0.0358} & 515.71 & \metrictablesecond{2.73} & 4.58K & 490.41 \\
\checkmark & 512 & 23.55 & 0.8457 & 0.1930 & 514.06 & 7.40 & 11.14K & 643.96 & 35.07 & \metrictablesecond{0.9775} & \metrictablebest{0.0354} & 514.24 & 4.34 & 4.61K & 488.84 \\
\checkmark & 1024 & \metrictablesecond{23.64} & \metrictablesecond{0.8476} & \metrictablesecond{0.1903} & 511.18 & 12.07 & 11.17K &  \metrictablesecond{635.00}  & \metrictablebest{35.11} & \metrictablebest{0.9776} & 0.0360 & 520.66 & 7.54 & \metrictablebest{4.56K} & \metrictablebest{485.39} \\
\checkmark & 2048 & \metrictablebest{23.66} & \metrictablebest{0.8487} & \metrictablebest{0.1901} & 505.60 & 21.40 & 11.10K & \metrictablebest{625.48} & 35.07 & 0.9771 & 0.0365 & \metrictablesecond{524.19} & 13.96 & \metrictablesecond{4.56K} & \metrictablesecond{486.10} \\
\checkmark & 4096 & 23.03 & 0.8316 & 0.2230 & 504.66 & 39.96 & \metrictablebest{10.56K} & 698.64 & \metrictablesecond{35.08} & 0.9769 & 0.0373 & \metrictablebest{524.41} & 26.80 & 4.57K & 492.66 \\
\bottomrule
\end{tabular}
}
\label{tab:nerf_ds_groups}
\end{table*}

\subsection{Ablation on TSS Hyperparameters}
\label{sec:appendix:tss}

Table~\ref{tab:appendix:tss} ablates the perturbation magnitude $\beta$ and annealing period $\tau$ for Temporal Sensitivity Sampling (TSS) on the NeRF-DS~\cite{yan2023nerf} dataset using the SpeeDe3DGS codebase. TSS improves pruning performance across all configurations, and its hyperparameters are not sensitive because the changes between timesteps in real-time video are small. We adopt the same $\beta{=}0.1$ and $\tau{=}20{,}000$ as DeformableGS~\cite{yang2023deformable3dgs}.

\begin{table*}
\caption{\textbf{Ablation on perturbation magnitude $\beta$ and annealing period $\tau$ on the NeRF-DS~\cite{yan2023nerf} dataset with our SpeeDe3DGS framework.} $\beta{=}-$ and $\tau{=}-$ indicate that TSS is not used. Each experiment is repeated three times and averaged to reduce training variance.}
\centering
\begin{tabular}{ccc|ccc}
\toprule
\textbf{Method} & $\beta$ & $\tau$ & PSNR~$\uparrow$ & SSIM~$\uparrow$ & LPIPS~$\downarrow$ \\
\midrule
\textbf{TSP} & - & - & 23.78 & 0.8507 & 0.1863 \\
\textbf{TSP + TSS} & 0.05 & 20K & 23.82 & 0.8513 & 0.1861 \\
\textbf{TSP + TSS} & 0.1 & 15K & 23.82 & 0.8510 & 0.1858 \\
\textbf{TSP + TSS} & 0.1 & 20K & 23.81 & 0.8515 & 0.1853 \\
\textbf{TSP + TSS} & 0.1 & 25K & 23.80 & 0.8517 & 0.1853 \\
\textbf{TSP + TSS} & 0.2 & 20K & 23.81 & 0.8514 & 0.1858 \\
\bottomrule
\end{tabular}
\label{tab:appendix:tss}
\end{table*}

\subsection{Ablation on Pruning Percentages}
\label{sec:appendix:percentages}

\begin{figure*}[h]
  \includegraphics[width=\linewidth]{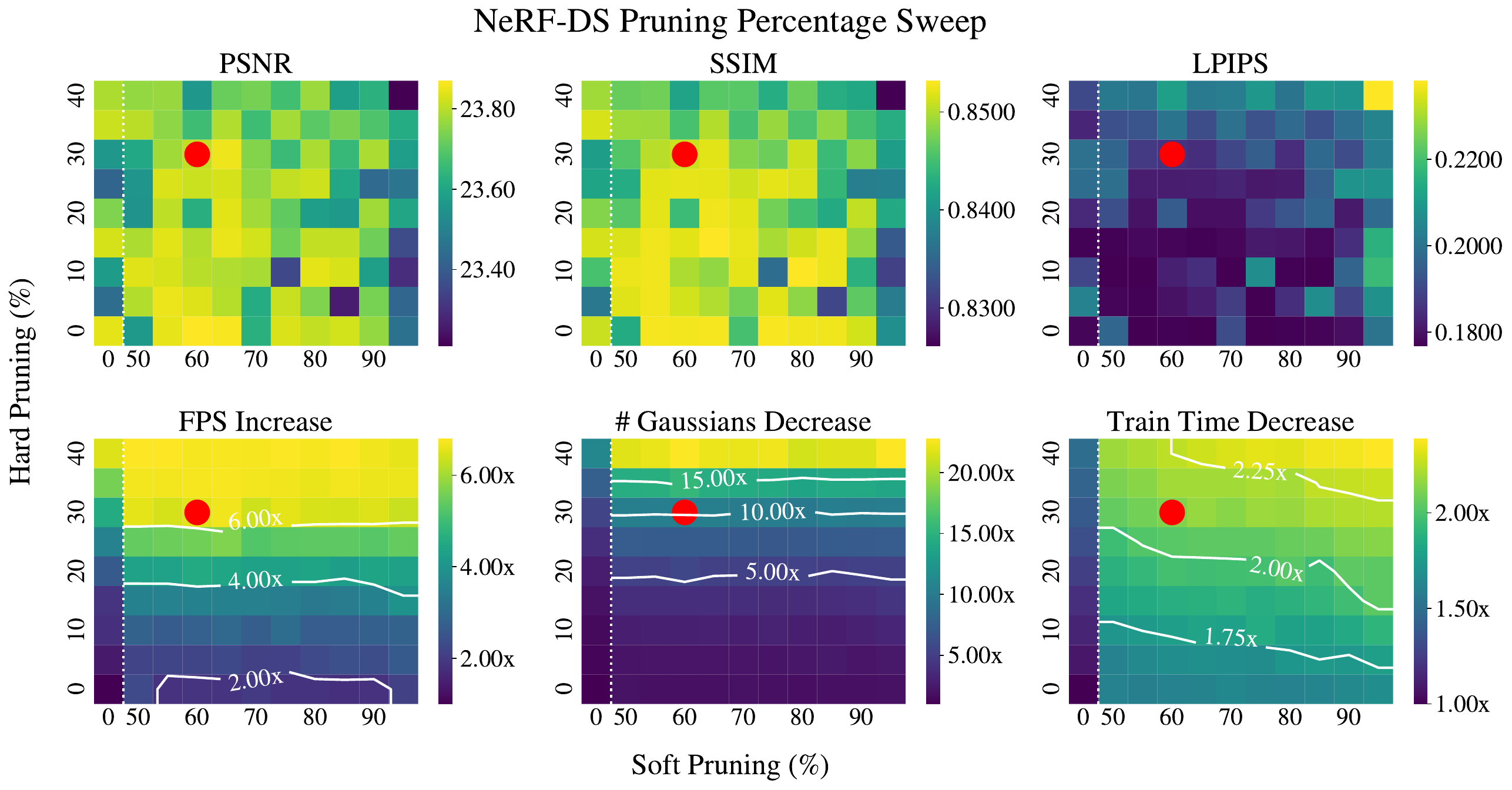}
  \includegraphics[width=\linewidth]{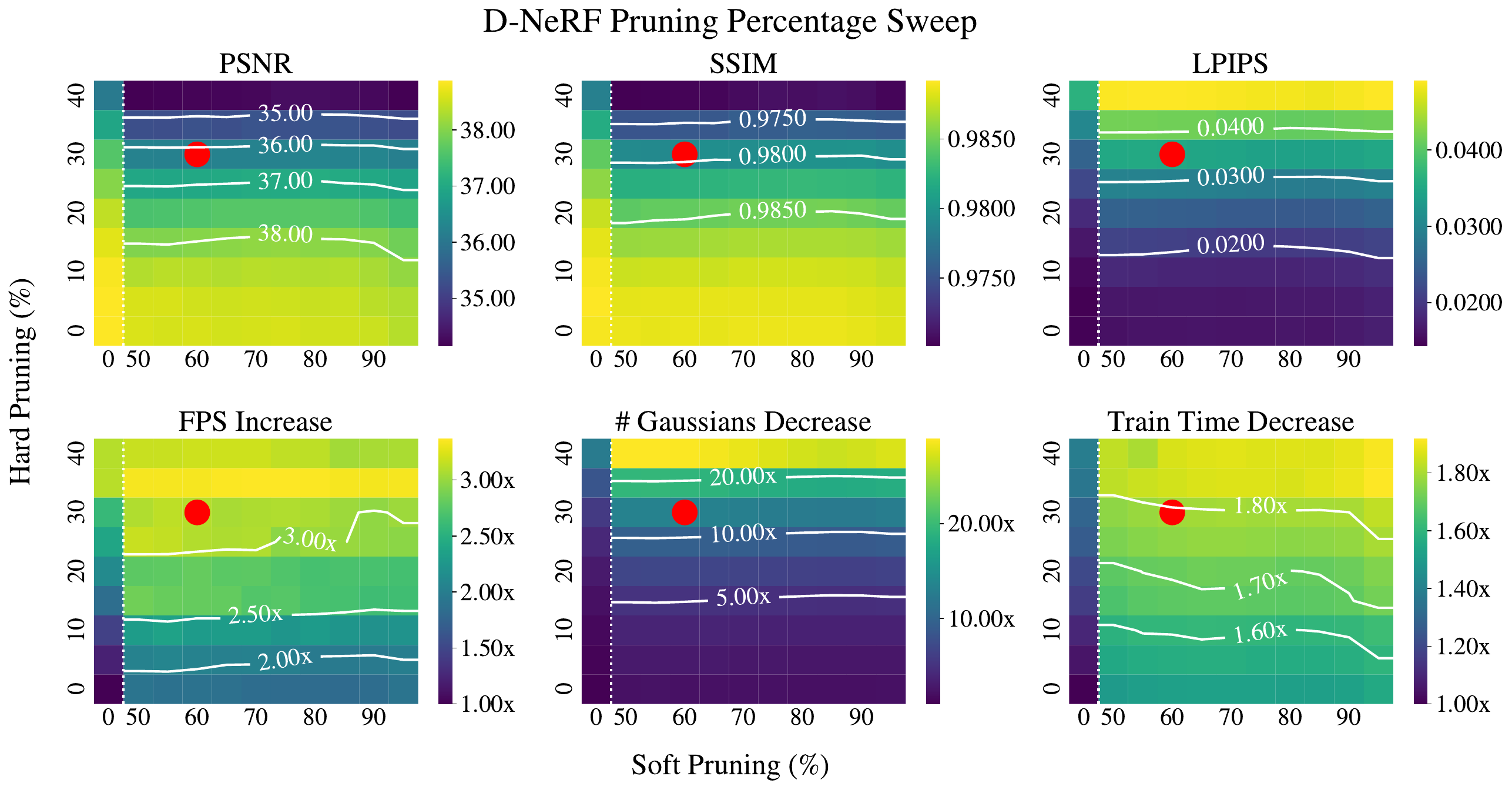}
  \vspace{-7mm}
\caption{\textbf{Ablation on pruning percentages with our SpeeDe3DGS framework.} We sweep soft (densification-stage) and hard (post-densification) pruning ratios in $5\%$ increments for the NeRF-DS~\cite{yan2023nerf} and D-NeRF~\cite{pumarola2020dnerf} datasets using the DeformableGS~\cite{yang2023deformable3dgs} codebase.
Each configuration is run three times without TSS or GroupFlow, and results are averaged across all runs.
($0\%$, $0\%$) corresponds to the unpruned baseline, while the first row and column show pruning in isolation.
The red dot marks our selected ($60\%$, $30\%$) soft–hard ratio.
FPS and Train Time improvements are measured on an RTX~A5000 GPU.}
  \label{fig:all_pruning}
\end{figure*}

Figure~\ref{fig:all_pruning} illustrates parameter sweeps over soft (densification-stage) and hard (post-densification) pruning percentages for the NeRF-DS~\cite{yan2023nerf} and D-NeRF~\cite{pumarola2020dnerf} datasets using the SpeeDe3DGS codebase.
Each configuration is evaluated in $5\%$ intervals and repeated three times to reduce variance, with results averaged across all runs.
All experiments are performed without Temporal Sensitivity Sampling (TSS) or GroupFlow to isolate the effect of Temporal Sensitivity Pruning (TSP) alone.
FPS and training time are collected on an RTX~A5000 GPU.
We empirically select the ($60\%$, $30\%$) soft–hard pruning ratio as it provides the best balance between rendering speed and visual quality.

\subsection{Per-Scene Metrics for SpeeDe3DGS}
\label{sec:appendix:scenes}

Tables~\ref{tab:app:psnr}, \ref{tab:app:ssim}, and \ref{tab:app:lpips} report per-scene PSNR, SSIM, and LPIPS, respectively, using our standalone SpeeDe3DGS codebase on NeRF-DS~\cite{yan2023nerf} and D-NeRF~\cite{pumarola2020dnerf}. Tables~\ref{tab:app:fps} and \ref{tab:app:train} provide per-scene rendering FPS and training time in seconds for NeRF-DS, D-NeRF, and the eight HyperNeRF~\cite{park2021hypernerf} scenes used in DeformableGS~\cite{yang2023deformable3dgs}. As in Appendix~\ref{sec:app:additional_speede3dgs}, rendering FPS is measured on an RTX~3090 GPU, while training time is measured on an RTX~A5000 GPU for NeRF-DS and D-NeRF and on an RTX~A6000 GPU for HyperNeRF.

\clearpage

\begin{table*}[h]

\caption{\textbf{PSNR~$\uparrow$ on each scene with our SpeeDe3DGS framework.} \emph{TSP}, \emph{TSS}, and \emph{GF} denote Temporal Sensitivity Pruning, Sampling, and GroupFlow, respectively. Each experiment is run three times and averaged to reduce training variance. The \metrictablebest{best} and \metrictablesecond{second best} results are highlighted.}
\centering
\resizebox{\linewidth}{!}{
\begin{tabular}{ccc|ccccccc|cccccccc}
\toprule
& & & \multicolumn{7}{c|}{{\small NeRF-DS~\cite{yan2023nerf}}} & \multicolumn{8}{c}{{\small D-NeRF~\cite{pumarola2020dnerf}}} \\
TSP & TSS & GF & \emph{as} & \emph{basin} & \emph{bell} & \emph{cup} & \emph{plate} & \emph{press} & \emph{sieve} & \emph{balls} & \emph{warrior} & \emph{hook} & \emph{jacks} & \emph{lego} & \emph{mutant} & \emph{standup} & \emph{trex} \\
\midrule
\multicolumn{3}{c|}{DeformableGS~\cite{yang2023deformable3dgs}} & 26.09 & 19.69 & 25.00 & \metrictablesecond{24.61} & 20.36 & 25.32 & 25.12 & \metrictablebest{40.92} & \metrictablebest{41.22} & \metrictablebest{36.68} & \metrictablebest{37.40} & \metrictablebest{32.52} & \metrictablebest{41.90} & \metrictablebest{43.44} & \metrictablebest{37.25} \\

\checkmark &  &  & \metrictablebest{26.15} & \metrictablesecond{19.54} & 25.20 & 24.56 & \metrictablebest{20.39} & \metrictablebest{25.49} & 25.08 & 38.24 & 39.47 & 34.18 & \metrictablesecond{35.54} & 30.58 & 37.02 & 39.08 & 35.44 \\

\checkmark & \checkmark &   &  \metrictablesecond{26.10} & \metrictablebest{19.67} & \metrictablebest{25.46} & \metrictablebest{24.63} & \metrictablesecond{20.36} & \metrictablesecond{25.47} & 24.87  & 38.32 & 39.44 & 34.16 & 35.43 & 30.47 & 36.96 & 39.11 & \metrictablesecond{35.62} \\
 
&  & \checkmark &  25.78 & 19.44 & 25.06 & 24.04 & 20.22 & 25.05 & \metrictablesecond{25.21} & \metrictablesecond{38.44} & \metrictablesecond{40.10} & \metrictablesecond{35.12} & 35.06 & \metrictablesecond{31.31} & \metrictablesecond{39.81} & \metrictablesecond{40.03} & 34.97 \\

\checkmark & \checkmark & \checkmark & 26.05 & 19.47 & \metrictablesecond{25.22} & 24.30 & 20.22 & 25.09 & \metrictablebest{25.25} & 36.68 & 38.86 & 33.24 & 34.04 & 30.05 & 36.32 & 37.51 & 33.85 \\
\bottomrule
\end{tabular}
}
\label{tab:app:psnr}
\end{table*}

\begin{table*}[h]

\caption{\textbf{SSIM~$\uparrow$ on each scene with our SpeeDe3DGS framework.}}
\centering
\resizebox{\linewidth}{!}{
\begin{tabular}{ccc|ccccccc|cccccccc}
\toprule
& & & \multicolumn{7}{c|}{{\small NeRF-DS~\cite{yan2023nerf}}} & \multicolumn{8}{c}{{\small D-NeRF~\cite{pumarola2020dnerf}}} \\
TSP & TSS & GF & \emph{as} & \emph{basin} & \emph{bell} & \emph{cup} & \emph{plate} & \emph{press} & \emph{sieve} & \emph{balls} & \emph{warrior} & \emph{hook} & \emph{jacks} & \emph{lego} & \emph{mutant} & \emph{standup} & \emph{trex} \\
\midrule
\multicolumn{3}{c|}{DeformableGS~\cite{yang2023deformable3dgs}} & 0.8781 & \metrictablesecond{0.7935} & 0.8424 & \metrictablebest{0.8895} & 0.8100 & 0.8616 & \metrictablebest{0.8712} & \metrictablebest{0.9955} & \metrictablebest{0.9866} & \metrictablebest{0.9850} & \metrictablebest{0.9892} & \metrictablebest{0.9766} & \metrictablebest{0.9941} & \metrictablebest{0.9939} & \metrictablebest{0.9925} \\

\checkmark &  &   & \metrictablesecond{0.8820} & 0.7912 & 0.8436 & 0.8881 & \metrictablebest{0.8126} & \metrictablebest{0.8636} & \metrictablesecond{0.8694} & 0.9925 & 0.9792 & 0.9721 & 0.9819 & 0.9598 & 0.9783 & 0.9849 & 0.9856 \\

\checkmark & \checkmark &   & \metrictablebest{0.8838} & \metrictablebest{0.7947} & \metrictablebest{0.8498} & \metrictablesecond{0.8886} & \metrictablesecond{0.8116} & \metrictablesecond{0.8622} & 0.8676 & 0.9925 & 0.9793 & 0.9719 & 0.9817 & 0.9591 & 0.9781 & 0.9849 & 0.9858 \\

&  & \checkmark &  0.8748 & 0.7837 & 0.8411 & 0.8790 & 0.8022 & 0.8562 & 0.8661 & \metrictablesecond{0.9932} & \metrictablesecond{0.9835} & \metrictablesecond{0.9813} & \metrictablesecond{0.9856} & \metrictablesecond{0.9727} & \metrictablesecond{0.9924} & \metrictablesecond{0.9907} & \metrictablesecond{0.9900} \\

\checkmark & \checkmark & \checkmark & 0.8799 & 0.7890 & \metrictablesecond{0.8466} & 0.8864 & 0.8080 & 0.8620 & 0.8688 & 0.9905 & 0.9771 & 0.9695 & 0.9790 & 0.9567 & 0.9777 & 0.9828 & 0.9836 \\
\bottomrule
\end{tabular}
}
\label{tab:app:ssim}
\end{table*}

\begin{table*}[h]
\caption{\textbf{LPIPS~$\downarrow$ on each scene with our SpeeDe3DGS framework.}}
\centering
\resizebox{\linewidth}{!}{
\begin{tabular}{ccc|ccccccc|cccccccc}
\toprule
& & & \multicolumn{7}{c|}{{\small NeRF-DS~\cite{yan2023nerf}}} & \multicolumn{8}{c}{{\small D-NeRF~\cite{pumarola2020dnerf}}} \\
TSP & TSS & GF & \emph{as} & \emph{basin} & \emph{bell} & \emph{cup} & \emph{plate} & \emph{press} & \emph{sieve} & \emph{balls} & \emph{warrior} & \emph{hook} & \emph{jacks} & \emph{lego} & \emph{mutant} & \emph{standup} & \emph{trex} \\
\midrule
\multicolumn{3}{c|}{DeformableGS~\cite{yang2023deformable3dgs}} & \metrictablebest{0.1861} & \metrictablebest{0.1856} & \metrictablebest{0.1611} & \metrictablebest{0.1553} & \metrictablebest{0.2231} & \metrictablebest{0.1914} & \metrictablebest{0.1475} & \metrictablebest{0.0088} & \metrictablebest{0.0262} & \metrictablebest{0.0174} & \metrictablebest{0.0144} & \metrictablebest{0.0212} & \metrictablebest{0.0069} & \metrictablebest{0.0088} & \metrictablebest{0.0107} \\

\checkmark &  &   & \metrictablesecond{0.1914} & 0.1974 & 0.1770 & 0.1602 & 0.2299 & 0.2004 & 0.1582 & 0.0177 & 0.0525 & 0.0436 & 0.0291 & 0.0481 & 0.0360 & 0.0269 & 0.0254 \\

\checkmark & \checkmark &   &  0.1922 & \metrictablesecond{0.1957} & \metrictablesecond{0.1628} & \metrictablesecond{0.1599} & \metrictablesecond{0.2293} & \metrictablesecond{0.1996} & 0.1596 & 0.0178 & 0.0523 & 0.0441 & 0.0292 & 0.0489 & 0.0361 & 0.0269 & 0.0251 \\
 
&  & \checkmark &  0.1961 & 0.2040 & 0.1654 & 0.1695 & 0.2337 & 0.1996 & \metrictablesecond{0.1564} & \metrictablesecond{0.0127} & \metrictablesecond{0.0309} & \metrictablesecond{0.0204} & \metrictablesecond{0.0172} & \metrictablesecond{0.0232} & \metrictablesecond{0.0088} & \metrictablesecond{0.0118} & \metrictablesecond{0.0126} \\

\checkmark & \checkmark & \checkmark & 0.1971 & 0.2052 & 0.1653 & 0.1646 & 0.2356 & 0.2020 & 0.1610 & 0.0216 & 0.0551 & 0.0446 & 0.0312 & 0.0491 & 0.0360 & 0.0285 & 0.0258 \\
\bottomrule
\end{tabular}
}
\label{tab:app:lpips}
\end{table*}

\begin{table*}[h]
\caption{\textbf{FPS~$\uparrow$ on each scene with our SpeeDe3DGS framework.}  Metrics are collected on an RTX~3090 GPU for consistency with MonoDyGauBench~\cite{liang2025monocular}.}
\centering
\resizebox{\linewidth}{!}{
\begin{tabular}{ccc|ccccccc|cccccccc|cccccccc}
\toprule
& & & \multicolumn{7}{c|}{{\small NeRF-DS~\cite{yan2023nerf}}} & \multicolumn{8}{c|}{{\small D-NeRF~\cite{pumarola2020dnerf}}} & \multicolumn{8}{c}{{\small HyperNeRF~\cite{park2021hypernerf}}} \\
TSP & TSS & GF & \emph{as} & \emph{basin} & \emph{bell} & \emph{cup} & \emph{plate} & \emph{press} & \emph{sieve} & \emph{balls} & \emph{warrior} & \emph{hook} & \emph{jacks} & \emph{lego} & \emph{mutant} & \emph{standup} & \emph{trex} & \emph{americano} & \emph{chicken} & \emph{cookie} &  \emph{espresso}& \emph{hand} & \emph{lemon} & \emph{printer} &  \emph{torchocolate} \\
\midrule
\multicolumn{3}{c|}{\multirow{2}{*}{DeformableGS~\cite{yang2023deformable3dgs}}} & 51.64 & 40.32 & 34.82 & 49.22 & 48.58 & 50.65 & 51.62 & 85.36 & 277.08 & 98.70 & 182.66 & 43.71 & 80.71 & 185.68 & 65.87 & 8.89 & 13.24 & 11.34 & 25.39 & 5.37 & 17.13 & 21.40 & 13.07 \\
 &  &  & (1.00$\times$) & (1.00$\times$) & (1.00$\times$) & (1.00$\times$) & (1.00$\times$) & (1.00$\times$) & (1.00$\times$) & (1.00$\times$) & (1.00$\times$) & (1.00$\times$) & (1.00$\times$) & (1.00$\times$) & (1.00$\times$) & (1.00$\times$) & (1.00$\times$) & (1.00$\times$) & (1.00$\times$) & (1.00$\times$) & (1.00$\times$) & (1.00$\times$) & (1.00$\times$) & (1.00$\times$) & (1.00$\times$) \\

\multirow{2}{*}{\checkmark} & \multirow{2}{*}{} & \multirow{2}{*}{}   & \metrictablesecond{426.84} & 346.75 & 280.81 & 377.27 & 388.79 & 331.68 & \metrictablesecond{421.80} & 427.86 & 480.22 & \metrictablesecond{392.53} & 441.17 & 353.65 & \metrictablesecond{393.94} & \metrictablesecond{440.04} & 450.47 & 123.26 & 135.82 & 121.11 & 199.73 & 60.66 & 159.31 & 193.08 & 135.90 \\
&  &  & \metrictablesecond{(8.27$\times$)} & (8.60$\times$) & (8.06$\times$) & (7.66$\times$) & (8.00$\times$) & (6.55$\times$) & \metrictablesecond{(8.17$\times$)} & (5.01$\times$) & (1.73$\times$) & \metrictablesecond{(3.98$\times$)} & (2.42$\times$) & (8.09$\times$) & \metrictablesecond{(4.88$\times$)} & \metrictablesecond{(2.37$\times$)} & (6.84$\times$) & (13.87$\times$) & (10.26$\times$) & (10.68$\times$) & (7.87$\times$) & (11.30$\times$) & (9.30$\times$) & (9.02$\times$) & (10.40$\times$) \\

\multirow{2}{*}{\checkmark} & \multirow{2}{*}{\checkmark} & \multirow{2}{*}{} &  423.39 & 330.80 & 277.22 & \metrictablesecond{421.34} & 416.87 & 328.87 & 379.39 & \metrictablesecond{429.47} & 480.19 & 391.86 & \metrictablesecond{444.19} & \metrictablesecond{359.19} & 393.63 & 436.93 & \metrictablesecond{455.66} & 114.45 & 130.83 & 119.11 & 188.44 & 59.33 & 155.84 & 190.59 & 126.97 \\
& & & (8.20$\times$) & (8.20$\times$) & (7.96$\times$) & \metrictablesecond{(8.56$\times$)} & (8.58$\times$) & (6.49$\times$) & (7.35$\times$) & \metrictablesecond{(5.03$\times$)} & (1.73$\times$) & (3.97$\times$) & \metrictablesecond{(2.43$\times$)} & \metrictablesecond{(8.22$\times$)} & (4.88$\times$) & (2.35$\times$) & \metrictablesecond{(6.92$\times$)} & (12.87$\times$) & (9.88$\times$) & (10.5$\times$) & (7.42$\times$) & (11.05$\times$) & (9.10$\times$) & (8.91$\times$) & (9.71$\times$) \\
 
\multirow{2}{*}{} & \multirow{2}{*}{} & \multirow{2}{*}{\checkmark} &  418.65 & \metrictablesecond{387.38} & \metrictablesecond{394.63} & 413.78 & \metrictablesecond{424.97} & \metrictablesecond{421.47} & 382.58 & 330.31 & \metrictablesecond{489.38} & 378.30 & 417.95 & 300.25 & 328.20 & 428.41 & 321.76 & \metrictablesecond{167.79} & \metrictablesecond{245.01} & \metrictablesecond{177.56} & \metrictablesecond{320.61} & \metrictablesecond{129.54} & \metrictablesecond{217.72} & \metrictablesecond{316.91} & \metrictablesecond{239.04} \\
& & & (8.11$\times$) & \metrictablesecond{(9.61$\times$)} & \metrictablesecond{(11.33$\times$)} & (8.41$\times$) & \metrictablesecond{(8.75$\times$)} & \metrictablesecond{(8.32$\times$)} & (7.41$\times$) & (3.87$\times$) & \metrictablesecond{(1.77$\times$)} & (3.83$\times$) & (2.29$\times$) & (6.87$\times$) & (4.07$\times$) & (2.31$\times$) & (4.88$\times$) & \metrictablesecond{(18.87$\times$)} & \metrictablesecond{(18.51$\times$)} & \metrictablesecond{(15.66$\times$)} & \metrictablesecond{(12.63$\times$)} & \metrictablesecond{(24.12$\times$)} & \metrictablesecond{(12.71$\times$)} & \metrictablesecond{(14.81$\times$)} & \metrictablesecond{(18.29$\times$)} \\

\multirow{2}{*}{\checkmark} & \multirow{2}{*}{\checkmark} & \multirow{2}{*}{\checkmark} & \metrictablebest{522.29} & \metrictablebest{510.42} & \metrictablebest{499.90} & \metrictablebest{510.30} & \metrictablebest{517.44} & \metrictablebest{500.52} & \metrictablebest{478.33} & \metrictablebest{484.77} & \metrictablebest{546.52} & \metrictablebest{527.00} & \metrictablebest{548.85} & \metrictablebest{492.38} & \metrictablebest{511.32} & \metrictablebest{546.11} & \metrictablebest{536.57} & \metrictablebest{408.62} & \metrictablebest{438.86} & \metrictablebest{405.26} & \metrictablebest{427.20} & \metrictablebest{408.48} & \metrictablebest{401.04} & \metrictablebest{451.23} & \metrictablebest{442.34} \\
 & & & \metrictablebest{(10.11$\times$)} & \metrictablebest{(12.66$\times$)}& \metrictablebest{(14.36$\times$)} & \metrictablebest{(10.37$\times$)} & \metrictablebest{(10.65$\times$)} & \metrictablebest{(9.88$\times$)} & \metrictablebest{(9.27$\times$)} & \metrictablebest{(5.68$\times$)} & \metrictablebest{(1.97$\times$)} & \metrictablebest{(5.34$\times$)} & \metrictablebest{(3.00$\times$)} & \metrictablebest{(11.26$\times$)} & \metrictablebest{(6.34$\times$)} & \metrictablebest{(2.94$\times$)} & \metrictablebest{(8.15$\times$)} & \metrictablebest{(45.96$\times$)} & \metrictablebest{(33.15$\times$)} & \metrictablebest{(35.74$\times$)} & \metrictablebest{(16.83$\times$)} & \metrictablebest{(76.07$\times$)} & \metrictablebest{(23.41$\times$)} & \metrictablebest{(21.09$\times$)} & \metrictablebest{(33.84$\times$)} \\
\bottomrule
\end{tabular}
}
\label{tab:app:fps}
\end{table*}

\begin{table*}[h]
\caption{\textbf{Training time~$\downarrow$ in seconds on each scene with our SpeeDe3DGS framework.} Metrics are collected with an RTX~A5000 GPU for NeRF-DS~\cite{yan2023nerf} and D-NeRF~\cite{pumarola2020dnerf} and an RTX~A6000 GPU for HyperNeRF~\cite{park2021hypernerf}; see Appendix~\ref{sec:app:additional_speede3dgs} for more details.}
\centering
\resizebox{\linewidth}{!}{
\begin{tabular}{ccc|ccccccc|cccccccc|cccccccc}
\toprule
& & & \multicolumn{7}{c|}{{\small NeRF-DS~\cite{yan2023nerf}}} & \multicolumn{8}{c|}{{\small D-NeRF~\cite{pumarola2020dnerf}}} & \multicolumn{8}{c}{{\small HyperNeRF~\cite{park2021hypernerf}}} \\
TSP & TSS & GF & \emph{as} & \emph{basin} & \emph{bell} & \emph{cup} & \emph{plate} & \emph{press} & \emph{sieve} & \emph{balls} & \emph{warrior} & \emph{hook} & \emph{jacks} & \emph{lego} & \emph{mutant} & \emph{standup} & \emph{trex} & \emph{americano} & \emph{chicken} & \emph{cookie} &  \emph{espresso}& \emph{hand} & \emph{lemom} & \emph{printer} &  \emph{torchocolate} \\
\midrule
\multicolumn{3}{c|}{\multirow{2}{*}{DeformableGS~\cite{yang2023deformable3dgs}}} & 1407.08 & 1692.53 & 1924.65 & 1467.75 & 1378.45 & 1404.10 & 1386.70 & 1090.20 & 490.14 & 888.91 & 604.36 & 1604.89 & 1021.78 & 580.32 & 1243.60 & 6746.79 & 4138.62 & 6145.04 & 3156.94 & 10118.35 & 4482.59 & 2660.25 & 4542.70 \\
 &  &  & (1.00$\times$) & (1.00$\times$) & (1.00$\times$) & (1.00$\times$) & (1.00$\times$) & (1.00$\times$) & (1.00$\times$) & (1.00$\times$) & (1.00$\times$) & (1.00$\times$) & (1.00$\times$) & (1.00$\times$) & (1.00$\times$) & (1.00$\times$) & (1.00$\times$) & (1.00$\times$) & (1.00$\times$) & (1.00$\times$) & (1.00$\times$) & (1.00$\times$) & (1.00$\times$) & (1.00$\times$) & (1.00$\times$) \\

\multirow{2}{*}{\checkmark} & \multirow{2}{*}{} & \multirow{2}{*}{}   & 723.34 & \metrictablesecond{776.01} & \metrictablesecond{940.58} & \metrictablesecond{748.42} & 626.04 & \metrictablesecond{683.67} & 714.11 & 579.25 & \metrictablesecond{381.99} & 523.37 & 412.22 & \metrictablesecond{741.95} & \metrictablesecond{540.69} & 401.63 & 615.06 & 2426.49 & \metrictablesecond{1432.92} & 2216.52 & 1383.00 & 3867.42 & \metrictablesecond{1744.81} & \metrictablesecond{1040.99} & 1826.96 \\
 &  &  & (1.95$\times$) & \metrictablesecond{(2.18$\times$)} & \metrictablesecond{(2.05$\times$)} & \metrictablesecond{(1.96$\times$)} & (2.20$\times$) & \metrictablesecond{(2.05$\times$)} & (1.94$\times$) & (1.88$\times$) & \metrictablesecond{(1.28$\times$)} & (1.7$\times$) & (1.47$\times$) & \metrictablesecond{(2.16$\times$)} & \metrictablesecond{(1.89$\times$)} & (1.44$\times$) & (2.02$\times$) & (2.78$\times$) & \metrictablesecond{(2.89$\times$)} & (2.77$\times$) & (2.28$\times$) & (2.62$\times$) & \metrictablesecond{(2.57$\times$)} & \metrictablesecond{(2.56$\times$)} & (2.49$\times$) \\

\multirow{2}{*}{\checkmark} & \multirow{2}{*}{\checkmark} & \multirow{2}{*}{}   &  \metrictablesecond{709.59} & 801.15 & 961.06 & 763.10 & \metrictablesecond{616.79} & 716.46 & \metrictablesecond{707.28} & \metrictablesecond{575.82} & 383.70 & \metrictablesecond{521.08} & \metrictablesecond{411.99} & 746.84 & 540.70 & \metrictablesecond{400.76} & \metrictablesecond{613.80} & \metrictablesecond{2409.60} & 1441.43 & \metrictablesecond{2185.75} & \metrictablesecond{1368.90} & \metrictablesecond{3646.07} & 1799.51 & 1072.80 & \metrictablesecond{1825.71} \\
 &  &  & \metrictablesecond{(1.98$\times$)} & (2.11$\times$) & (2.0$\times$) & (1.92$\times$) & \metrictablesecond{(2.23$\times$)} & (1.96$\times$) & \metrictablesecond{(1.96$\times$)} & \metrictablesecond{(1.89$\times$)} & (1.28$\times$) & \metrictablesecond{(1.71$\times$)} & \metrictablesecond{(1.47$\times$)} & (2.15$\times$) & (1.89$\times$) & \metrictablesecond{(1.45$\times$)} & \metrictablesecond{(2.03$\times$)} & \metrictablesecond{(2.80$\times$)} & (2.87$\times$) & \metrictablesecond{(2.81$\times$)} & \metrictablesecond{(2.31$\times$)} & \metrictablesecond{(2.78$\times$)} & (2.49$\times$) & (2.48$\times$) & \metrictablesecond{(2.49$\times$)} \\
 
\multirow{2}{*}{} & \multirow{2}{*}{} & \multirow{2}{*}{\checkmark} &  758.47 & 843.14 & 1011.61 & 824.64 & 729.46 & 800.62 & 819.34 & 798.95 & 403.14 & 577.25 & 498.60 & 936.23 & 660.42 & 445.82 & 770.69 & 3283.08 & 1794.64 & 2971.33 & 1644.66 & 4539.15 & 2179.02 & 1252.77 & 2378.86 \\
 &  & & (1.86$\times$) & (2.01$\times$) & (1.9$\times$) & (1.78$\times$) & (1.89$\times$) & (1.75$\times$) & (1.69$\times$) & (1.36$\times$) & (1.22$\times$) & (1.54$\times$) & (1.21$\times$) & (1.71$\times$) & (1.55$\times$) & (1.3$\times$) & (1.61$\times$) & (2.06$\times$) & (2.31$\times$) & (2.07$\times$) & (1.92$\times$) & (2.23$\times$) & (2.06$\times$) & (2.12$\times$) & (1.91$\times$) \\

\multirow{2}{*}{\checkmark} & \multirow{2}{*}{\checkmark} & \multirow{2}{*}{\checkmark} & \metrictablebest{600.07} & \metrictablebest{640.83} & \metrictablebest{754.56} & \metrictablebest{638.72} & \metrictablebest{531.30} & \metrictablebest{592.14} & \metrictablebest{620.75} & \metrictablebest{590.77} & \metrictablebest{371.01} & \metrictablebest{449.02} & \metrictablebest{431.48} & \metrictablebest{625.99} & \metrictablebest{482.93} & \metrictablebest{381.81} & \metrictablebest{555.79} & \metrictablebest{1756.25} & \metrictablebest{976.64} & \metrictablebest{1612.27} & \metrictablebest{1119.53} & \metrictablebest{2300.66} & \metrictablebest{1321.41} & \metrictablebest{810.68} & \metrictablebest{1317.33} \\
& & & \metrictablebest{(2.34$\times$)} & \metrictablebest{(2.64$\times$)} & \metrictablebest{(2.55$\times$)} & \metrictablebest{(2.3$\times$)} & \metrictablebest{(2.59$\times$)} & \metrictablebest{(2.37$\times$)} & \metrictablebest{(2.23$\times$)} & \metrictablebest{(1.85$\times$)} & \metrictablebest{(1.32$\times$)} & \metrictablebest{(1.98$\times$)} & \metrictablebest{(1.4$\times$)} & \metrictablebest{(2.56$\times$)} & \metrictablebest{(2.12$\times$)} & \metrictablebest{(1.52$\times$)} & \metrictablebest{(2.24$\times$)} & \metrictablebest{(3.84$\times$)} & \metrictablebest{(4.24$\times$)} & \metrictablebest{(3.81$\times$)} & \metrictablebest{(2.82$\times$)} & \metrictablebest{(4.40$\times$)} & \metrictablebest{(3.39$\times$)} & \metrictablebest{(3.28$\times$)} & \metrictablebest{(3.45$\times$)} \\
\bottomrule
\end{tabular}
}
\label{tab:app:train}
\end{table*}

\onecolumn

\begin{figure*}[t]
  \includegraphics[width=\linewidth]{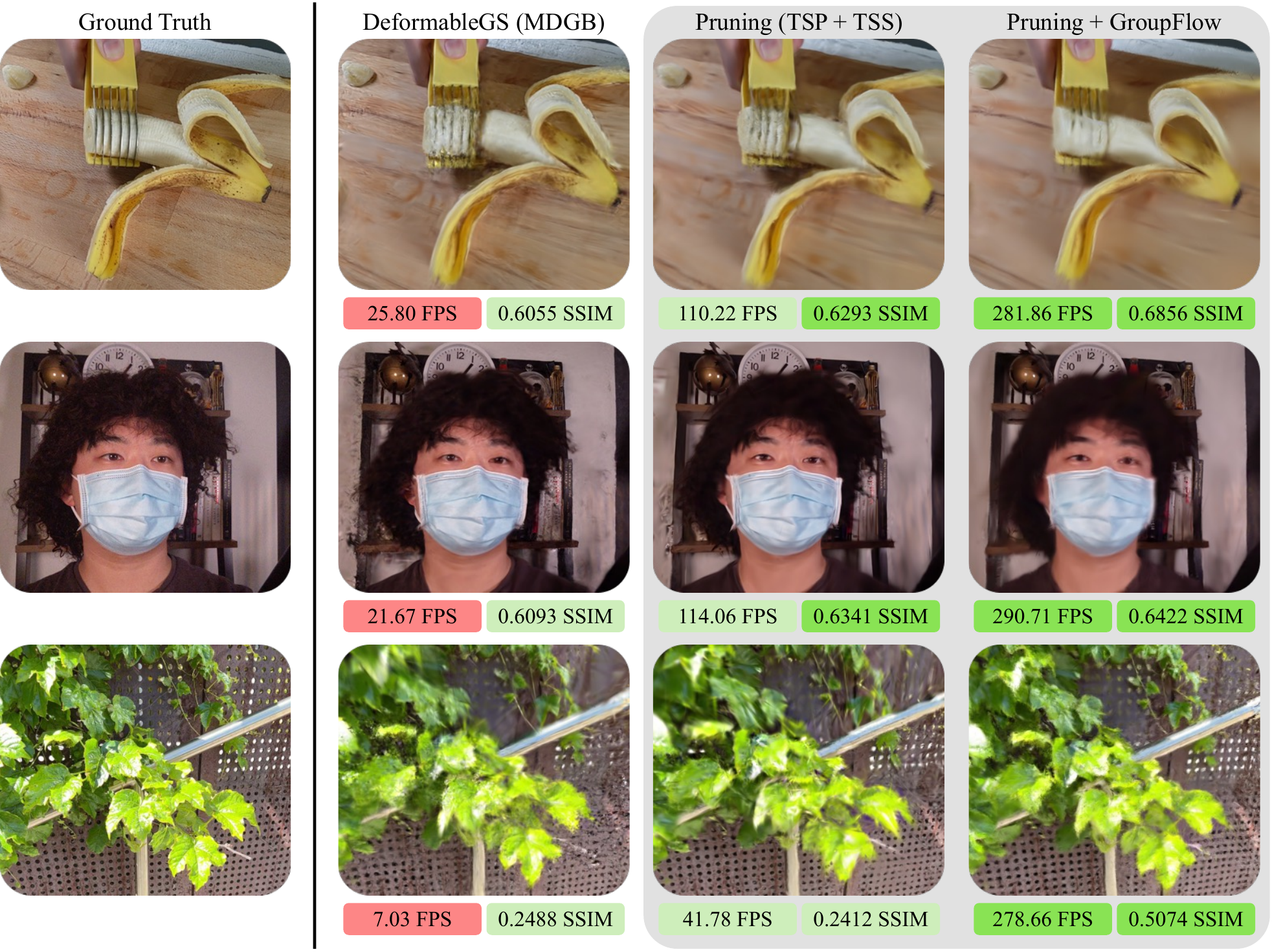}
  \vspace{-6mm}
\caption{\textbf{Visual comparison of the baseline DeformableGS~\cite{yang2023deformable3dgs} and our SpeeDe3DGS methods on MonoDyGauBench (MDGB)~\cite{liang2025monocular}.} The baseline examples are reproduced through retraining, as the original MonoDyGauBench models are not available. For all other visualizations, we use our standalone SpeeDe3DGS codebase, which produces consistently higher FPS and image quality than the MDGB wrapper. Top: \emph{slice-banana} from HyperNeRF~\cite{park2021hypernerf}. Middle: \emph{curls} from Nerfies~\cite{park2021nerfies}. Bottom: \emph{creeper} from iPhone~\cite{gao2022dynamic}.}
  \label{fig:mgdb_qualitative}
  \vspace{-3mm}
\end{figure*}

\begin{multicols}{2}

\subsection{Qualitative Results on MonoDyGauBench}
\label{sec:app:mdgb_qualitative}
Figure~\ref{fig:mgdb_qualitative} presents qualitative comparisons between the DeformableGS~\cite{yang2023deformable3dgs} baseline and our SpeeDe3DGS methods, using representative scenes from HyperNeRF~\cite{park2021hypernerf}, Nerfies~\cite{park2021nerfies}, and iPhone~\cite{gao2022dynamic} within the MonoDyGauBench framework~\cite{liang2025monocular}. Pruning (TSP+TSS) and GroupFlow significantly accelerate rendering while maintaining comparable or better visual fidelity than the baseline. As discussed in Appendix~\ref{sec:app:monodygaubench}, grouped SE(3) motion distillation also acts as a regularizer in real-world scenes -- enhancing temporal coherence, mitigating drift under noisy or unstable camera trajectories, and improving overall image quality. This effect is especially pronounced on the \emph{creeper} scene from iPhone, where the SSIM of the Pruning+GroupFlow model is twice that of both the baseline and Pruning-only variants. Note that our standalone SpeeDe3DGS codebase achieves superior rendering speed and visual fidelity compared to the standardized MonoDyGauBench framework.

\columnbreak

\subsection{Baseline and Motion Model Limitations}
\label{sec:appendix:limitations}

Our methods inherit the limitations of the dynamic Gaussian Splatting frameworks into which they are integrated. In particular, highly deformable scenes remain challenging for current approaches. MonoDyGauBench~\cite{liang2025monocular} finds that the baseline methods struggle with non-rigid motion and noisy camera poses, so our approach faces similar limitations.

Nevertheless, Sections~\ref{sec:experiments:monodygaubench} and Appendices~\ref{sec:app:monodygaubench} and~\ref{sec:app:groups} demonstrate that GroupFlow can introduce regularizing effects that improve temporal stability in such scenarios. In practice, globally non-rigid scenes consist of clusters of small Gaussians whose motion is locally rigid due to cloning and splitting during densification. Increasing the number of groups $J$ enables progressively finer motion modeling; in the extreme case $J{=}N$, GroupFlow distills per-Gaussian neural motion into efficient SE(3) transformations. Adaptive strategies that refine groups based on motion variation could further improve modeling of complex non-rigid motion.

\end{multicols}

\end{document}